\documentclass[nofootinbib,superscriptaddress, showpacs,preprintnumbers, referee, twocolumn, nofootinbibt]{revtex4-2}
\usepackage{eurosym}
\usepackage{amsmath}
\usepackage{bm}
\usepackage{amsfonts}
\usepackage{amssymb}
\usepackage{float}
\usepackage{graphicx}
\usepackage{subfig}
\usepackage{caption}
\usepackage{comment}
\usepackage{multirow}
\usepackage[export]{adjustbox}
\graphicspath{ {figures} }
\usepackage{xcolor}
\usepackage{hyperref}
\usepackage{tikz}
\usetikzlibrary{shapes.geometric, arrows}
\usepackage[figure, ruled]{algorithm2e}
\usepackage{booktabs}

\def\be{\begin{equation}}
\def\ee{\end{equation}}

\begin{document}
	
	\title{Big Bang Nucleosynthesis constraints on space-time noncommutativity}
	\author{Teodora M. Matei}
	\email{teodora.maria.matei@stud.ubbcluj.ro}
	\affiliation{Faculty of Physics, Babe\c s-Bolyai University, 1 Kog\u alniceanu Street,
		Cluj-Napoca, 400084, Romania,}
	\affiliation{Astronomical Observatory, 19 Cire\c silor Street,
		Cluj-Napoca, 400487, Romania}
	\author{Cristian A. Croitoru}
	\email{croitoru.lu.cristian@student.utcluj.ro}
	\affiliation{Department of Computer Science, Technical University of Cluj-Napoca, G. Baritiu Street 26-28, Cluj-Napoca 400027, Romania}
	\author{Tiberiu Harko}
	\email{tiberiu.harko@aira.astro.ro}
	\affiliation{Faculty of Physics, Babe\c s-Bolyai University, 1 Kog\u alniceanu Street,
		Cluj-Napoca, 400084, Romania,}
	\affiliation{Astronomical Observatory, 19 Cire\c silor Street,
		Cluj-Napoca, 400487, Romania}
	
	\begin{abstract}
		We consider the implications of the modified dispersion relations, due to the noncommutativity of the spacetime, for a photon gas filling the early Universe in the framework of the Big Bang Nucleosynthesis (BBN) processes, during the period of light elements formation.  We consider three types of deformations present in the dispersion relations for the radiation gas, from which we obtain the low temperature corrections to the energy density and pressure. The cosmological implications of the modified equations of state in the BBN era are explored in detail for all radiation models. The effects induced on the nucleosynthesis process by spacetime noncommutativity are investigated by evaluating the abundances of relic nuclei (Hydrogen, Deuterium, Helium-3, Helium-4, and Lithium-7).  The primordial mass fraction estimates and their deviations due to changes in the freezing temperature impose an upper limit on the  energy density of the deformed photon gas, which follows from the modified Friedmann equations. The deviations from the standard energy density of the radiative plasma are therefore constrained by the abundances of the Helium-4 nuclei. Upper limits on the free parameters of the spacetime noncommutativity are obtained via a numerical analysis performed using the \texttt{PRyMordial} software package. The primordial abundances of the light elements are obtained by evaluating the thermonuclear reaction rates for the considered noncommutative spacetime models. An MCMC (Markov Chain Monte Carlo) analysis allows to obtain restrictions on the free parameters of the modified dispersion relations. The numerical and statistical approach is implemented in the python code \texttt{PRyNCe}, available on GitHub.
	\end{abstract}
	
	\maketitle
	\tableofcontents
	
	\section{Introduction}
	
	The concepts of noncommutative spacetime and geometry originated as theoretical proposals for solving two fundamental problems in physics. The first is the incompatibility between the best presently known theory of gravity -- general relativity -- and quantum physics. The second open class of problems is represented by the challenges particle physics and cosmology presently face, such as the nonlocality, singularity,  dark energy, and dark matter problems, respectively. A possible solution to these problems can be found by assuming the finite nature of the two fundamental concepts of physics, length and time,  on the Planck scales.  This assumption allows us to consistently and systematically understand many phenomena at both microscopic and macroscopic (astrophysical and cosmological) levels \cite{N1,N2,N3,N4,N5}. The finite nature of the Planck length and time points towards the existence of the noncommutative spacetime, and towards the necessity of introducing  noncommutative algebra and geometry in the description of natural phenomena. The noncommutative mathematical and physical concepts can be further generalized to the idea of the noncommutative phase space, representing  a generalization of the Heisenberg algebra of quantum mechanics \cite{N6,N7}.  Moreover, the systematic use of noncommutative geometric and algebraic 
	concepts and techniques could lead to a solution of the singularity and nonlocality problems in quantum gravity, quantum cosmology, and particle physics.
	
	Several mathematical techniques have been proposed and investigated in order to implement the idea of noncommutative phase space \cite{N6}.  Some of these are represented by the canonical formulation, the Moyal product, Bopp shift,  path integration, the Weyl--Wigner phase space \cite{N8,N9}, and the Seiberg--Witten map \cite{N10,N11,N12}. The noncommutative
	phase space generalizes in a consistent way the uncertainty principle, and smears the phase space, thus leading to new physical effects arising from the noncommutative
	nature of the spacetime and phase space. Interesting physical implications of the noncommutative spacetime, such as the Berry phase \cite{N13}, the Aharonov--Bohm effect \cite{N14,N15,N16,N17,N18}, quantum Hall and spin Hall effects \cite{N19,N19a}, and magnetic monopoles \cite{N20} have been extensively investigated. The three-dimensional noncommutative relations of the position and momentum operators have been generalized to four dimensions in \cite{N20a}. Using the Seiberg--Witten map, the Heisenberg representation of these noncommutative algebras was given,  and the noncommutative parameters associated with the Planck constant, Planck length and cosmological constant were provided. The propagation of photons in the Moyal space half-filled with a static and homogeneous electric field was considered in \cite{N20b}. The electromagnetic fluctuations on top of this step-like background were analyzed, and the localization of photons and the possibility of photon production by strong electric fields were considered. 
	
	Unfortunately, up to now, no direct experimental evidence to detect the presence of a noncommutative geometry is known. The main difficulty in directly observing physical effects caused
	by spacetime or phase space noncommutativity is due to the fact that these effects are too weak, and generally they are relevant on the Planck
	scale only. However, there are some observable physical effects that could lead to constraints on the structure of the noncommutative spacetime. In \cite{N21} and \cite{N21a} the noncommutative
	field theory and the Lorentz violation were studied and an upper bound of the noncommutative parameter, $\eta \leq \left(10\; {\rm TeV}\right)^{-2}$  was obtained. 
	A method for exploring the spatial noncommutativity of the scattering differential cross-section by using the Aharanov--Bohm effect was proposed in \cite{N22}. The proposal involves  particle physics experiments in the energy ranges between 200 and 300 GeV for $\eta \leq \left(10\; {\rm TeV}\right)^{-2}$, and for the typical order of magnitude of the cross-section for neutrino events. 
	
	The existence of a persistent current in a nanoscale ring in the presence of an external magnetic field along the ring axis was investigated in \cite{N23}. Two observable physical quantities were introduced that allow to probe the signal coming from the noncommutative phase space. A value-independent criterion to prove the existence of the noncommutative phase space was also introduced.
	
	The standard theoretical approach to cosmology, namely the Big Bang theory,  is constructed with the help of three fundamental observational principles. These are the Hubble expansion of the Universe, which follows from the observations of the galactic redshift, the Cosmic Microwave Background Radiation, studied by several satellite missions, like, such as the Planck satellite \cite{Planck}, and the Big Bang Nucleosynthesis (BBN), respectively. Hence, the Big Bang theory is supported by a large number of observational data. In
	particular, the BBN theory plays a crucial role in understanding the physics of the early Universe, since it successfully predicts the cosmological abundances of light elements such as D, $^{3,4}$He and $^{6,7}$Li, respectively. The formation of the light elements took place after the Big Bang, during a period of rapid expansion of the Universe, which allowed for the formation of the lightest elements only. However, during the BBN phase, the formation of radioactive or unstable isotopes, such as $^3$H and $^{7,8}$Be, also took place. The decay of these unstable isotopes led to the increase of the concentration of stable elements. 
	
	The expansion of the Universe caused the temperature and the density of the Universe to drop, and under these physical conditions the formation of elements heavier than beryllium could not take place. Hence, light elements became the dominant form of baryonic matter in the very early Universe.
	
	The BBN period began three minutes after the formation of the Universe, lasted around seventeen minutes, and
	ended when the Universe was about twenty minutes old.
	Before the BBN period, the temperature of the Universe decreased from around 
	$10^{16}$ GeV  to 1 MeV, which is the temperature at which the density of the nucleon component allowed the formation of stable nuclei via nuclear reactions \cite{Copi, Fields}. Hence,  the reduced nucleon concentration and the rapid expansion of the Universe led to the formation of light nuclei only, like  
	hydrogen, which is the most abundant element in the Universe, helium, the second most common element, as well as an isotope of lithium. The Big Bang Nucleosynthesis lasted until the temperature of the Universe dropped below several keV due to cosmic expansion, which caused primordial nuclear reactions to stop.
	
	The standard model of cosmology, as well as the BBN theory, are based on the standard model of particle physics, which includes, among other physical characteristics, three families of
	neutrinos. During the BBN the formation of the light nuclei depended on several important physical parameters such as expansion rate,  nucleon density, temperature, neutrino-antineutrino asymmetry and neutrino abundance, respectively. There is a very good agreement between the theoretical predictions of the BBN theory of the primordial abundances of D and $^{3,4}$He, and the observational results. However, the BBN theory is facing serious challenges since the theoretical predictions for the abundance of $^{6,7}$Li show a significant difference with respect to the observations \cite{Fields-2011}.
	
	Therefore, BBN represents an important theoretical approach that allows to test the predictions of the cosmological scenarios on  the conditions
	under which the primordial nucleosynthesis of elements formed \cite{Copi, Fields, Fields-2011, Steigman-2004,   Sepico, Lambiase-2005, Mangano2005,lesgourgues, Coc, Steigman-2007, Molaro2008,
		Steigman-2008, Steigman-2012,Coc2017, Cybrut, Husdal-2016,obs2, Benstein2, Pitrou, Kohri, F, Hsyu, Aver,obs1,yeh,tau}.
	
	In particular, BBN data can be used to provide in-depth tests of modified gravity theories. Constraints on the $f(R,T)$ gravity theory were obtained and analyzed in \cite{S1}. The effects of the nonextensive thermostatistics were investigated by considering primordial helium abundance in \cite{Torres}. Constraint from BBN on massive modified gravity theories were obtained in \cite{Lambiase-2012}. The first order corrections to the energy densities and weak interaction rates  were obtained, and they were used to compute the deviation in the primordial helium abundance. In \cite{Barrow}  BBN constraints on the exponent of the Barrow entropy have been obtained.   The constraints on the parameters of the bimetric gravity were obtained using BBN data in \cite{M1}. Viable $f(T)$ teleparallel cosmological models (power law, exponential and square-root exponential) were analysed for finding upper limits satisfying the BBN data on primordial abundance of $^4$He in \cite{Capozziello}. The implications of higher-order modified gravity theories were investigated by using BBN in \cite{M2}. By using observational data coming from the Big Bang Nucleosynthesis and the matter–antimatter asymmetry in the baryogenesis era,  constraints on the bumblebee timelike vector field were obtained in \cite{M4a}. Limits on the size of the Lorentz violation, and the rate of the time evolution of the background bumblebee field were also determined. The $f \left(T ,T_G\right)$ modified theory of gravity, were $T$ is the torsion scalar, and $T_G$ is the teleparallel equivalent of the Gauss–Bonnet term, was constrained, by using the BBN data, in \cite{M3}.  The implications for the BBN of the 
	$f(Q)$ and $f(Q,T)$ type theories of gravity were investigated in \cite{M4,M5,M6}. An analysis of the BBN implications of the $f(T,\tau)$ gravity theory was performed in \cite{S2}. Big Bang Nuclesynthesis constraints on modified gravity theories in the presence of a Weylian boundary were considered in \cite{M7}. 
	
	It is the main goal of the present paper to investigate the constraints imposed on the noncommutative nature of  spacetime, which can be obtained by comparing the theoretical models with the BBN data. Hence, we consider the evolution of the very early Universe, as well as during the BBN phase,  by assuming that the radiation source takes the form of a deformed photon gas. When the energy scales of the cosmological processes are high enough, quantum effects on the structure of  space-time must also be considered. One of the essential physical and cosmological implications of space-time noncommutativity is the modification of the dispersion relation, which has important implications on the statistical and thermodynamic properties 
	of the elementary particles, including photons \cite{disp1}.
	
	As a first step in our study, we investigate the implications of the spacetime noncommutative structure on the statistical mechanics of radiation, which therefore affects cosmological expansion. Furthermore, we use BBN data to obtain constraints on the free parameters of three different models of noncommutative spacetimes, each of them characterized by its own modified dispersion
	relations. The first modified commutation relation we consider  has the spacetime commutators given by $\left[\hat{x}^i, \hat{t}\right] = i\lambda \hat{x}^i$, $\left[\hat{x}^i, \hat{x}^j\right] = 0$.  The astrophysical implications of these modified commutation relations were studied in \cite{disp1}. The second and the third models we will consider within the BBN framework are described by the modified Heisenberg relations, $\left[\hat{x}^i, \hat{p}^j\right] = i\hbar \delta^{ij} (1 + \beta_0 p^2)$, $\left[\hat{p}^i, \hat{p}^j\right] = 0$,  and $\left[\hat{x}^i, \hat{p}^j\right] = i\hbar \left[\delta^{ij} - \alpha_0(p \delta^{ij} + \frac{p^i p^j}{p}) + \alpha_0^2(p^2 \delta^{ij} + 3p^i p^j) \right]$, $ \left[\hat{p}^i, \hat{p}^j\right] = 0$, respectively, where $\alpha _0$ and $\beta _0$ are constants \cite{disp2}. Generally, the modified Heisenberg relations can be associated with the noncommutative spacetime geometry by the use of the Jacobi identity. The cosmological implications of the modified equations of state for radiation have been considered in the modified Friedmann equation from loop quantum cosmology in \cite{lqc}. The evolution equations of basic cosmological parameters (scale factor, Hubble function, radiation temperature, and deceleration parameter) were investigated by numerical integrating a deformed radiation gas model in the early stages of the Universe. In all models, the evolution of the Universe indicates the presence of a (nonsingular) bounce, which describes the transition from a contracting to an expanding phase.
	
	As a next step in our present analysis, we consider the statistical properties of the deformed photon gas, described by the generalized  uncertainty and dispersion relations. Our theoretical approach is based on the general form of the partition function $Z$ for a thermodynamic system, from which one can obtain the general expressions of the energy density and pressure of the deformed photon gas, satisfying the  modified dispersion relations considered. From the general formalism, we obtain the equations of state for the photon gas in the cosmologically relevant limit of low temperatures and densities, a range that is relevant for the BBN processes. Even in the low temperature limit, the existence of noncommutativity in the spacetime structure
	leads to important modifications in the equation of state of radiation. This will lead to an altered  dependence of the photon gas energy density and pressure on temperature as compared to the  expressions of the black body radiation in the standard formulation of quantum mechanics and statistical physics. 
	
	Once the modified equations of state for radiation are known, we can constrain the  noncommutative spacetime models by using the observational values of the  Helium-4 abundance, which are obtained from the primordial mass fraction estimate. In this respect, we use the correlated deviation of the freeze-out temperature, following an approach discussed in \cite{Lambiase-2012} and \cite{Capozziello}, respectively. The freeze-out temperature is the temperature at which the neutron and proton interconversion processes via weak interactions stop due to the temperature decrease of the Universe, as a result of cosmic expansion. Estimates of the primordial mass fraction and its deviation are considered in \cite{ Hsyu, Aver,obs1} and \cite{pdg}, respectively. Therefore, the contribution to the photon energy density coming from the noncommutative nature of space-time can be constrained by obtaining an upper limit coming from the $\delta Y_p/Y_p$ fraction.
	
	Moreover, we will use the deformed energy density of the photon gas, together with its constrained contribution within the  \texttt{PRyMordial} software package \cite{primordial}, which allows us to analyze the three  cosmological models in the full BBN framework. The light nuclei abundances are obtained using the thermonuclear reaction rates available in the NACRE II database \cite{nacre2}, together with the evaluation of the Friedmann cosmological equations describing the evolution of primordial plasma during the BBN phase. By using an MCMC (Markov Chain Monte Carlo) analysis, a full set of constraints for the parameters of the deformed photon gas are obtained.
	
	The present paper is organized as follows. We review the fundamentals of the statistical mechanics and consider the noncommutative spacetime geometries in Section~\ref{generaldisp}. The equations of state of the deformed photon gas are also obtained in the low temperature limit.  The theoretical formalism for the description of the BBN processes in the presence of spacetime noncommutativity is presented in Section~\ref{BBN}. The statistical methodology for the comparison of the theoretical predictions with the BBN data, as well as the results of the numerical fits and of the MCMC analysis are presented in Section~\ref{Stat}. We discuss and conclude our results in Section~\ref{Concl}. 
	
	\section{Dispersion relations in noncommutative spacetimes}\label{generaldisp}
	
	In noncommutative quantum mechanics, the classical phase space coordinates and momenta are replaced by quantum operators that generalize the non-commutation relations of the standard quantum mechanical description. Hence, in the noncommutative phase space, the commutation relations are given by \cite{N23}
	\be
	\left[\hat{x}^\mu,\hat{x}^\nu\right]=i\theta ^{\mu \nu}, \left[\hat{p}^\mu,\hat{p}^\nu\right]=i\eta ^{\mu \nu}, \left[\hat{x}^\mu,\hat{p}^\nu\right]=i\Delta ^{\mu \nu},
	\ee
	where $\theta ^{\mu \nu}$ and $\eta ^{\mu \nu}$ are the noncommutative strength parameters.  The quantisation of space in the early Universe brings modifications to the behaviour of relativistic particles near the Planck scale, also having an impact on the evolution of the matter sector in the expanding Universe. In the case of massless particles such as photons, these effects manifest  by inducing deformed dispersion relations, which influence the physical properties of the electromagnetic radiation. In this Section, we introduce three models of noncommutative spacetimes with their corresponding non-commutative relations, and we obtain the modifications  induced by the modified spacetime geometry to the energy density and pressure of a photon gas.
	
	\subsection{General statistical framework}
	
	We consider a photon gas with two polarization states and no chemical potential. The grand-canonical partition function $Z$ is given by \cite{lqc}
	\begin{equation}
		Z = \prod_{\vec{k}} \sum_{n_{\vec{k}} = 0}^\infty e^{-\beta \hbar \omega(k) n_{\vec{k}}}
		= \prod_{\vec{k}} \frac{1}{1 - e^{-\beta \hbar \omega(k)}},
		\label{Z_general}
	\end{equation}
	where $\beta = 1/(k_B T)$, $k_B$ is Boltzmann's constant,  and $\omega(k)$ is the model-dependent dispersion relation. The mean energy is obtained as
	\begin{equation}
		U = \sum_{\vec{k}} \hbar \omega(k) \langle n_{\vec{k}} \rangle,
	\end{equation}
	where the occupation number follows the Bose-Einstein distribution for massless particles
	\begin{equation}
		\langle n_{\vec{k}} \rangle = \frac{1}{e^{\beta \hbar \omega(k)} - 1}.
	\end{equation}
	
	Summing over polarizations and integrating over momenta, we obtain
	\begin{align}
		U &= 2 \sum_{\vec{k}} \frac{\hbar \omega(k)}{e^{\beta \hbar \omega(k)} - 1}
		= \frac{2V}{(2\pi)^3} \int d^3k \, \frac{\hbar \omega(k)}{e^{\beta \hbar \omega(k)} - 1} \notag \\
		&= \frac{2V}{(2\pi)^3} \int_0^\infty dk \, 4\pi k^2 \frac{\hbar \omega(k)}{e^{\beta \hbar \omega(k)} - 1},
		\label{U_integral}
	\end{align}
	where the factor of 2 appears from the two polarization states of the photon.
	
	The energy density of the photon gas is thus given by
	\begin{equation}
		\rho = \frac{U}{V} = \frac{\hbar}{\pi^2} \int_0^\infty dk \, k^2 \frac{\omega(k)}{e^{\beta \hbar \omega(k)} - 1}.
		\label{rho_general}
	\end{equation}
	
	We further make a change of variables from $k$ to $\omega$.  For a continuous dispersion relation $\omega(k)$, we have
	\begin{equation*}
		k = k(\omega),  dk = \frac{dk}{d\omega} d\omega,
	\end{equation*}
	which allows us to write the energy density as
	\begin{equation}
		\rho = \frac{\hbar}{\pi^2} \int_0^\infty d\omega \, \frac{k^2(\omega)}{e^{\beta \hbar \omega} - 1} \omega \, \frac{dk}{d\omega}.
		\label{rho_change_var}
	\end{equation}
	
	The pressure is determined by the use of the thermodynamic relation
	\begin{equation}
		P = \frac{1}{\beta} \frac{\partial \ln Z}{\partial V},
	\end{equation}
%	By using Eq.~\eqref{Z_general}, we obtain first
%	\begin{align}
%		\ln Z &= - \sum_{\vec{k}, \vec{e}} \ln\left(1 - e^{-\beta \hbar \omega(k)}\right)
%		= -2 \sum_{\vec{k}} \ln\left(1 - e^{-\beta \hbar \omega(k)}\right),
%	\end{align}
	and by using Eq.~(\ref{Z_general}) we find 
	\begin{eqnarray*}
		P&=& -\frac{1}{\beta} \partial_V \cdot 2 \frac{V}{(2\pi)^3} \int d^3k \ln(1 - e^{-\beta\hbar \omega}) \notag \\
		&=& -\frac{1}{\beta \pi^2} \int_{0}^{\infty} dk \; k^2 \ln(1 - e^{-\beta\hbar \omega}) \notag \\
		&=& -\frac{1}{\beta \pi^2} \int_{0}^{\infty} dk \; d(k^3/3) \ln(1 - e^{-\beta\hbar \omega})   \\
		&=& \frac{1}{\beta \pi^2} \int_{0}^{\infty} dk \; \frac{k^3}{3} \frac{1}{e^{\beta\hbar \omega} - 1}\left(\beta\hbar \frac{d\omega}{dk}\right)   \notag \\
		&&-\frac{1}{\beta \pi^2} \int_{0}^{\infty} dk \; d\left(\frac{k^3 \ln(1 - e^{-\beta\hbar \omega})}{3}\right),
	\end{eqnarray*}
	where we accounted for the two polarization states of the photon.
	
	The second term is rather small and will be neglected. Therefore the pressure of the photon gas takes the form
	\begin{equation}
		P \simeq \frac{\hbar}{3\pi^2} \int_0^\infty dk \, \frac{k^3}{e^{\beta \hbar \omega(k)} - 1} \frac{d\omega}{dk}.
	\end{equation}
	
	By changing variables from $k$ to $\omega$, this becomes
	\begin{equation}
		P = \frac{\hbar}{3\pi^2} \int_0^\infty d\omega \, \frac{k^3(\omega)}{e^{\beta \hbar \omega} - 1}.
		\label{P_general}
	\end{equation}
	
	In the following, after specifying the form of the deformed dispersion relation $\omega =\omega(k)$ for three non-commutative models, we evaluate the thermodynamic quantities by using the general relations given in Eq.~\eqref{rho_change_var} and \eqref{P_general}, respectively. We focus on the low temperature regime, which is the one significant in the Big Bang Nucleosynthesis era.  Within this approximation  the non-commutative corrections can be expressed analytically as small deviations from the standard Stefan-Boltzmann law.
	
	\subsection{Model I: $\left[\hat{x}^i, \hat{t}\right] = i\lambda \hat{x}^i$, $\left[\hat{x}^i, \hat{x}^j\right] = 0$}\label{model1}
	
	In the first model, we are considering  the spacetime commutation relations given by
	\begin{equation}
		\left[\hat{x}^i, \hat{t}\right] = i\lambda \hat{x}^i, \; \left[\hat{x}^i, \hat{x}^j\right] = 0,
	\end{equation}
	where  $\lambda$ is the deformation parameter. The modified dispersion relation becomes  \cite{disp1, lqc, disp2},
	\begin{equation}
		\omega(k) = kc(1 + \lambda \hbar \omega).
	\end{equation}
	or in terms of $k$,
	\begin{equation}
		k(\omega) = \frac{\omega}{c(1 + \lambda \hbar \omega)}.
	\end{equation}
	By further differentiation, the following relation can be obtained
	\begin{equation}
		\frac{dk}{d\omega} = \frac{1}{c(1 + \lambda \hbar \omega)^2}.
	\end{equation}
	
	We now compute the energy density using Eq.~\eqref{rho_change_var}
	\begin{align}
		\rho_{(I)} &= \frac{\hbar}{\pi^2} \int_0^\infty d\omega \, \frac{k^2(\omega)}{e^{\beta \hbar \omega} - 1} \, \omega \, \frac{dk}{d\omega} \notag \\
		&= \frac{\hbar}{\pi^2} \int_0^\infty d\omega \, \frac{1}{(1 + \lambda \hbar \omega)^4} \cdot \frac{\omega^3}{c^3} \cdot \frac{1}{e^{\beta \hbar \omega} - 1}.
	\end{align}
	By changing the integration variable to $x = \beta \hbar \omega$, we obtain
	\begin{equation}
		\rho_{(I)} = \frac{(k_B T)^4}{\pi^2 c^3 \hbar^3} \int_0^\infty dx \, \frac{x^3}{(1 + x \lambda k_B T)^4 (e^x - 1)}.
	\end{equation}
	
	In the low temperature limit $\lambda k_B T \ll 1$, by using the following mathematical identity
	\begin{equation}
		\int_{0}^{\infty} \frac{x^{n-1}}{e^{ax} - 1}dx = \frac{\Gamma(n)\zeta(n)}{a^n}, \text{ Re } n>0, \text{ Re } a>0, \label{gamma}
	\end{equation}
	where $\zeta (n)$ is the Riemann zeta function, we obtain the energy density of the photon gas as  
	\begin{align}
		\rho_{(I)} &\simeq \frac{(k_B T)^4}{\pi^2 c^3 \hbar^3} \int_0^\infty dx \left(1 - 4x \lambda k_B T\right) \frac{x^3}{e^x - 1} \notag \\
		&= \frac{(k_B T)^4}{\pi^2 c^3 \hbar^3} \left( \frac{\pi^4}{15} - 96 \zeta(5) \lambda k_B T \right).
	\end{align}
	
	To compute the pressure, we use Eq.~\eqref{P_general}, where the momentum $k(\omega)$ is given by
	\begin{equation}
		k(\omega) = \frac{\omega}{c(1 + \lambda \hbar \omega)}.
	\end{equation}
	
	Hence the pressure becomes
	\begin{equation}
		P_{(I)} = \frac{\hbar}{3\pi^2} \int_0^\infty d\omega \, \frac{\omega^3}{(1 + \lambda \hbar \omega)^3 c^3} \cdot \frac{1}{e^{\beta \hbar \omega} - 1}.
	\end{equation}
	
	With the substitution $x = \beta \hbar \omega$, we obtain
	\begin{equation}
		P_{(I)} = \frac{(k_B T)^4}{3\pi^2 c^3 \hbar^3} \int_0^\infty dx \, \frac{x^3}{(1 + x \lambda k_B T)^3 (e^x - 1)}.
	\end{equation}
	
	By expanding the integrand in same low temperature regime, and after making use of Eq. \eqref{gamma}, we obtain
	\begin{align}
		P_{(I)} &\simeq \frac{(k_B T)^4}{3\pi^2 c^3 \hbar^3} \left( \frac{\pi^4}{15} - 72 \zeta(5) \lambda k_B T \right).
	\end{align}
	
	It can be observed that in the low temperature regime, the spacetime noncommutative effects introduce a correction to the standard thermodynamic quantities linear in \(\lambda k_B T\), resulting in a small deviation from the standard energy density and pressure of radiation.
	
	\subsection{Model II: $\left[\hat{x}^i, \hat{p}^j\right] = i\hbar \delta^{ij} (1 + \beta p^2)$, $\left[\hat{p}^i, \hat{p}^j\right] = 0$ }\label{model2}
	
	In the second model we are considering that the position operator takes the form $\hat{x}_i = x_{0i}$, and the momentum operator is given by $\hat{p}_i = p_{0i}(1+\beta_0 p^2)$, respectively. The modified Heisenberg relations become \cite{lqc}
	\begin{equation}
		\left[\hat{x}^i, \hat{p}^j\right] = i\hbar \delta^{ij} (1 + \beta p^2),
	\end{equation} 
	and
	\begin{equation}
		\left[\hat{p}^i, \hat{p}^j\right] = 0,
	\end{equation}
	where $\beta_0$ is the deformation parameter, and $p = \sqrt{g_{ij}p^{0i}p^{0j}}$. Therefore, in the Minkowski spacetime
	\begin{eqnarray*}
		p_a p^a &=& g_{a b} p^{a} p^{b} = - g_{00}(p^0) + g_{ij}p^{0i}p^{0j}(1 + \beta_0 p^2)^2 \notag \\
		&\simeq& -(p^0)^2 + p^2 + 2\beta_0 p^2 p^2 = -m^2c^2  
	\end{eqnarray*}
	which yields
	\begin{equation}
		(p^0)^2 = m^2c^2 + p^2(1 - 2\beta_0 p^2).
	\end{equation}
	
	With this quadratic correction applied, the modified dispersion relation becomes \cite{disp2}
	\begin{equation}
		\omega(k) = kc \sqrt{1 - 2\beta_0 \hbar^2 \omega^2}.
	\end{equation}
	Solving for $k(\omega)$ gives
	\begin{equation}
		k = \frac{\omega}{c \sqrt{1 - 2\beta_0 \hbar^2 \omega^2}}, \qquad
		\frac{dk}{d\omega} = \frac{1}{c(1 - 2\beta_0 \hbar^2 \omega^2)^{3/2}}.
	\end{equation}
	
	Using Eq.~\eqref{rho_change_var}, the photon gas energy density becomes
	\begin{align}
		\rho_{(II)} &= \frac{\hbar}{\pi^2} \int_0^\infty d\omega \, \frac{\omega^3}{c^3 (1 - 2\beta_0 \hbar^2 \omega^2)^{5/2}} \cdot \frac{1}{e^{\beta \hbar \omega} - 1}.
	\end{align}
	By making use of the same substitution $x = \beta \hbar \omega$, the energy density of the radiation becomes
	\begin{equation}
		\rho_{(II)} = \frac{(k_B T)^4}{\pi^2 c^3 \hbar^3} \int_0^\infty dx \, \frac{x^3}{(1 - 2 \beta_0 (k_B T)^2 x^2)^{5/2}(e^x - 1)}.
	\end{equation}
	
	By further expanding the denominator, we obtain in the low temperature regime $\beta_0(\kappa_B T)^2 \ll 1$ the following energy density
	\begin{align}
		\rho_{(II)} &\simeq \frac{(k_B T)^4}{\pi^2 c^3 \hbar^3} \left( \frac{\pi^4}{15} + 600 \zeta(6) \beta_0 (k_B T)^2 \right).
	\end{align}
	
	The pressure follows from Eq.~\eqref{P_general}, and it is obtained in a general form as
	\begin{equation}
		P_{(II)} = \frac{(k_B T)^4}{3\pi^2 c^3 \hbar^3} \int_0^\infty dx \, \frac{x^3}{(1 - 2 \beta_0 (k_B T)^2 x^2)^{3/2}(e^x - 1)}.
	\end{equation}
	By using again a series expansion, the pressure for the deformed photon gas becomes
	\begin{align}
		P_{(II)} &\simeq \frac{(k_B T)^4}{3\pi^2 c^3 \hbar^3} \left( \frac{\pi^4}{15} + 360 \zeta(6) \beta_0 (k_B T)^2 \right).
	\end{align}
	
	The thermodynamic corrections in this model scale with $\beta_0 (k_B T)^2$, leading to a quadratic increase in temperature of both energy density and pressure of the photon gas as compared to the standard radiation case.
	
	\subsection{Model III: $\left[\hat{x}^i, \hat{p}^j\right] = i\hbar \left(\delta^{ij} - \alpha_0(p \delta^{ij} + \frac{p^i p^j}{p}) + \alpha_0^2(p^2 \delta^{ij} + 3p^i p^j) \right)$, $ \left[\hat{p}^i, \hat{p}^j\right] = 0$ }\label{model3}
	
	In the third model we consider the position and the momentum operators to be constructed as $\hat{x}_i = x_{i0}$ and $\hat{p}_i = p_{0i}(1 - \alpha_0 p + 2 \alpha_0^2 p^2) $, which give the following generalized Heisenberg relations \cite{lqc}
	\begin{equation}
		\left[\hat{x}^i, \hat{p}^j\right] = i\hbar \left[\delta^{ij} - \alpha_0\left(p \delta^{ij} + \frac{p^i p^j}{p}\right) + \alpha_0^2(p^2 \delta^{ij} + 3p^i p^j) \right],
	\end{equation} 
	and
	\begin{equation}
		\left[\hat{p}^i, \hat{p}^j\right] = 0,
	\end{equation}
	where $\alpha_0$ is the dimensionless deformation parameter and $p = \sqrt{g_{ij}p^{0i}p^{0j}}$. Therefore, in the Minkowski spacetime
	\begin{eqnarray*}
		p_a p^a &=& g_{a b} p^{a} p^{b} = - g_{00}(p^0) + g_{ij}p^{0i}p^{0j}(1 - \alpha_0 p + 2\alpha_0^2 p^2)^2 \notag \\
		&\simeq& -(p^0)^2 + p^2 - 2\alpha_0 p^3 = -m^2c^2, 
	\end{eqnarray*}
	since we consider only the first order approximation. Then the following relation is immediately obtained
	\begin{equation}
		\left(p^0\right)^2 = m^2c^2 + p^2\left(1 - 2\alpha_0 p\right).
	\end{equation}
	The dispersion relation for this model is
	\begin{equation}
		\omega(k) = kc \sqrt{1 - 2\alpha_0 \hbar \omega},
	\end{equation}
	which allows us to express $k(\omega)$ as
	\begin{equation}
		k = \frac{\omega}{c \sqrt{1 - 2\alpha_0 \hbar \omega}}, \qquad
		\frac{dk}{d\omega} = \frac{1 - \alpha_0 \hbar \omega}{c (1 - 2\alpha_0 \hbar \omega)^{3/2}}.
	\end{equation}
	
	The energy density for the deformed photon gas becomes
	\begin{align}
		\rho_{(III)} &= \frac{\hbar}{\pi^2 c^3} \int_0^\infty d\omega \, \frac{1 - \alpha_0 \hbar \omega}{(1 - 2\alpha_0 \hbar \omega)^{5/2}} \cdot \frac{\omega^3}{e^{\beta \hbar \omega} - 1} \\
		&\approx \frac{\hbar}{\pi^2 c^3} \int_0^\infty d\omega \, \frac{1}{(1 - 2\alpha_0 \hbar \omega)^{5/2}} \frac{\omega^3}{e^{\beta \hbar \omega}}.
	\end{align}
	By applying the variable change $x = \beta \hbar \omega$, we obtain
	\begin{equation}
		\rho_{(III)} = \frac{(k_B T)^4}{\pi^2 c^3 \hbar^3} \int_0^\infty dx  \frac{1}{(1 - 2\alpha_0 k_B T x)^{5/2}}  \frac{x^3}{e^x - 1}.
	\end{equation}
	
	Expanding the integral in the \(\alpha_0 k_B T \ll 1\) regime, the energy density of the photon gas is obtained as
	\begin{align}
		\rho_{(III)} &\simeq \frac{(k_B T)^4}{\pi^2 c^3 \hbar^3} \left( \frac{\pi^4}{15} + 120 \zeta(5) \alpha_0 k_B T \right).
	\end{align}
	
	The pressure has the general expression
	\begin{equation}
		P_{(III)} = \frac{(k_B T)^4}{3\pi^2 c^3 \hbar^3} \int_0^\infty dx \, \frac{x^3}{(1 - 2x\alpha_0 k_B T)^{3/2}(e^x - 1)},
	\end{equation}
	which in the low temperature limit gives
	\begin{align}
		P_{(III)} &\simeq \frac{(k_B T)^4}{3\pi^2 c^3 \hbar^3} \left( \frac{\pi^4}{15} + 72 \zeta(5) \alpha_0 k_B T \right).
	\end{align}
	
	Hence, in this model spacetime noncommutativity adds a new correction term proportional to the temperature into the thermodynamic quantities describing the photon gas.

	\section{Big Bang Nucleosynthesis in noncommutative Spacetimes}\label{BBN}
	
	Big Bang Nucleosynthesis describes the formation of light nuclei in the temperature range $T \sim 1~\mathrm{MeV}$ and $T \sim 0.1~\mathrm{MeV}$. Within the first three minutes after the Big Bang, the Universe consisted of dense and hot plasma, composed mostly of photons, leptons and baryon, with all these particle species in thermal equilibrium. In the present  Section we first review some of the fundamental aspects of BBN theory, including the decoupling of the weakly interacting particles, and the evolution of the scale factor of the Universe, 
	quantities that are extremely sensitive to the total radiation content of the early Universe. Then, we will present the basic equations and the numerical approach for studying the BBN processes in the presence of a deformed photon gas. The nucleosynthesis results obtained can impose upper limits to the contributions of the deformation parameters of the photon gas considered in the noncommutative spacetime models. 
	
	\subsection{Overview of the BBN process and light element formation}
	
	The theoretical basis of the BBN was established by Gamow, Alpher, and Bethe \cite{Gamow}, who first analyzed the nucleosynthesis processes in the early Universe by considering that the primordial plasma underwent a series of nuclear reactions that resulted in the formation of light elements. Later on, Peebles \cite{Peebles1966} introduced time-dependent neutron evolution and incorporated Coulomb barrier effects, allowing for a more accurate description of the synthesis of heavier elements. The weak processes are incorporated as
	\begin{equation}
		n + \nu_e \leftrightarrow p + e^-, \qquad n \rightarrow p + e^- + \overline{\nu}_e.
	\end{equation}
	
	These reactions maintained chemical equilibrium as long as the weak reaction rate $\Gamma_{n \leftrightarrows p} \sim G_F^2 T^5$ exceeded the Hubble expansion rate $H \sim T^2/M_{\mathrm{Pl}}$. As the Universe expanded and cooled down, the equilibrium broke down when $\Gamma(T) \leq H(T)$, leading to particle decoupling from the relativistic plasma, as weak interactions become inefficient. Following this freeze-out period, which occurred at approximately $T_f \simeq 0.5$--$0.6~\mathrm{MeV}$, non-relativistic particle species left behind a relic abundance determined by their concentration at decoupling \cite{Benstein}.
	
	The weak interaction rate can be written  explicitly as \cite{Benstein}
	\begin{equation}
		\Lambda(T) = 4A T^3 (4!T^2 + 2 \cdot 3! Q T + 2! Q^2),
	\end{equation}
	where \( Q = m_n - m_p = 1.29 \times 10^{-3}~\mathrm{GeV} \) and \( A = 1.02 \times 10^{-11}~\mathrm{GeV}^{-4} \). At high temperatures, this reduces to the leading term \cite{Benstein, Lambiase-2012}
	\begin{equation}
		\Lambda(T) \simeq q T^5, \; q = 9.6 \times 10^{-10}~\mathrm{GeV}^{-4}.
	\end{equation}
	
	The particle decoupling has a direct impact on the synthesis of light elements, as the neutron abundance at freeze-out determined the total Helium-4 production through the baryon-to-photon ratio, given as
	\begin{equation}
		\eta_B \equiv \frac{n_b}{n_\gamma},
	\end{equation}
	which controls the efficiency of nuclear reactions. Higher values of $ \eta_B $ lead to an increase in nuclear reaction rates, and, therefore, increased helium production and reduced deuterium abundance. 
	
	Since $ \eta_B $ is sensitive to the expansion rate and entropy evolution of the Universe, any modification to the Friedmann equations, such as those induced by noncommutative radiation thermodynamics, can produce a shift in the predicted light element abundances.
	
	During the radiation-dominated epoch, the energy density is given by
	\begin{equation}
		\rho_r = \frac{\pi^2}{30} g_* T^4, \qquad g_* \sim 10, \label{enden}
	\end{equation}
	and the expansion rate follows from the first Friedmann equation
	\begin{equation}
		H(T) = \left( \frac{\pi^2 g_*}{90} \right)^{1/2} \frac{T^2}{M_{\mathrm{Pl}}}. \label{Hubble}
	\end{equation}
	
	By setting $ \Lambda(T_f) = H(T_f) $, the freeze-out temperature can be obtained as
	\begin{equation}
		T_f = \left( \frac{\pi^2 g_*}{90 q^2 M_{\mathrm{Pl}}^2} \right)^{1/6} \sim 0.5~\mathrm{MeV}.
	\end{equation}
	
	At freeze-out, the neutron-to-proton ratio becomes
	\begin{equation}
		\left(\frac{n}{p}\right)_{T_f} \simeq e^{-Q/T_f},
	\end{equation}
	with $Q = m_n - m_p \approx 1.293~\mathrm{MeV}$. The neutron-to-proton ratio freezes at $n/p \simeq 1/6$. This quantity decays further through beta-decay processes to $1/7$, before nucleosynthesis begins. The Helium-4 mass fraction is then approximated by
	\begin{equation}
		Y_p = \lambda \frac{2(n/p)}{1 + (n/p)} \approx 0.25,
	\end{equation}
	where $ \lambda = \exp[-(t_n - t_f)/\tau_n] $ accounts for neutron decay, with $ \tau_n = 870.2 \pm 15.8~\mathrm{s} $ \cite{tau}. Variations in $T_f$ lead to a change in the helium abundance \cite{Capozziello},
	\begin{equation}
		\delta Y_p = Y_p \left[ \left( 1 - \frac{Y_p}{2\lambda} \right) \ln \left( \frac{2\lambda}{Y_p} - 1 \right) - \frac{2t_f}{\tau_n} \right] \frac{\delta T_f}{T_f},
	\end{equation}
	and therefore from the Helium-4 abundance we obtain the constraint on the freeze-out temperature as
	\begin{equation}
		\left| \frac{\delta T_f}{T_f} \right| < 4.7 \times 10^{-4}. \label{TfBound}
	\end{equation}

	\subsection{Nucleosynthesis in the presence of a deformed photon gas}
	
	In the standard cosmological model, the total radiation energy density before electron-positron annihilation is given by
	\begin{equation}
		\rho_r = \rho_\gamma + \rho_e + 3\rho_\nu = \frac{43}{8} \rho_\gamma,
	\end{equation}
	and it includes photons, relativistic electrons-positrons pairs, and three neutrino species. After $e^\pm$ annihilation, and incomplete neutrino decoupling, the radiation content is parameterized by the effective number of neutrino species $N_{\mathrm{eff}}$, so that
	\begin{equation}
		\rho_r = \rho_\gamma \left[ 1 + \frac{7}{8} \left( \frac{4}{11} \right)^{4/3} N_{\mathrm{eff}} \right].
	\end{equation}
	The Standard Model predicts $N_{\mathrm{eff}}^{\mathrm{SM}} = 3.046$, a result that is obtained by including non-instantaneous neutrino decoupling and finite-temperature QED corrections \cite{Mangano2005}.
	
	Any deviation from the standard model radiation content can be described by an additional contribution $\rho_X$, so that
	\begin{equation}
		\rho_r^{\mathrm{eff}} = \rho_r + \rho_X, \qquad \rho_X = \Delta N_\nu \, \rho_\nu,
	\end{equation}
	where
	\begin{equation}
		\Delta N_\nu = N_{\mathrm{eff}} - 3.046, \qquad \rho_\nu = \frac{7}{8} \left( \frac{T_\nu}{T_\gamma} \right)^4 \rho_\gamma.
	\end{equation}
	Current CMB and BBN observations constrain this deviation to $\Delta N_\nu < 0.18$ at $2\sigma$ \cite{yeh}, placing bounds on any new physics contribution to the effective energy density.
	
	The associated neutrino energy density parameter is given by
	\begin{equation}
		\Omega_\nu h^2 = \frac{\sum m_\nu}{93.14~\mathrm{eV}},
	\end{equation}
	which becomes relevant at later cosmological epochs.
	
	Within the noncommutative spacetime models, the photon dispersion relation is modified, leading to additional contributions to the thermodynamic quantities of the radiation background. The total energy density $\rho$ and pressure $P$ deviate from their standard expressions through quantum corrections, which are specific to each model considered in Section \ref{generaldisp}.
	
	In the following we consider a spatially flat and isotropic FLRW Universe, with line element given by
	\begin{equation}
		ds^2 = -c^2 dt^2 + a^2(t)\left(dx^2 + dy^2 + dz^2\right),
	\end{equation}
	where $a(t)$ is the scale factor. We also introduce an important observational quantity, the Hubble function, defined as $H(t)=\dot{a}/a$, where a dot denotes the derivative with respect to the cosmological time. 	In the FLRW metric the evolution of the Universe is described by the Friedmann and the continuity equations, given by 
	\begin{equation}
		H^2 = \frac{8\pi G}{3} \rho, \quad \dot{\rho} + 3H(\rho + P) = 0.
	\end{equation}
	
	The thermodynamic quantities of the photon gas were determined in the previous Section for each non-commutative model, and they are summarized below.
	
	\paragraph{Model I: $\omega(k) = kc(1 + \lambda \hbar \omega)$}
	\begin{align}
		\rho_{(I)} &= \frac{(k_B T)^4}{\pi^2 c^3 \hbar^3} \left( \frac{\pi^4}{15} - 96\zeta(5) \lambda k_B T \right), \\
		P_{(I)} &= \frac{(k_B T)^4}{3\pi^2 c^3 \hbar^3} \left( \frac{\pi^4}{15} - 72\zeta(5) \lambda k_B T \right).
	\end{align}
	
	\paragraph{Model II: $\omega(k) = kc \sqrt{1 - 2\beta_0 \hbar^2 \omega^2}$)}
	\begin{align}
		\rho_{(II)} &= \frac{(k_B T)^4}{\pi^2 c^3 \hbar^3} \left( \frac{\pi^4}{15} + 600\zeta(6)\beta_0 (k_B T)^2 \right), \\
		P_{(II)} &= \frac{(k_B T)^4}{3\pi^2 c^3 \hbar^3} \left( \frac{\pi^4}{15} + 360\zeta(6)\beta_0 (k_B T)^2 \right).
	\end{align}
	
	\paragraph{Model III: $\omega(k) = kc \sqrt{1 - 2\alpha_0 \hbar \omega}$}
	\begin{align}
		\rho_{(III)} &= \frac{(k_B T)^4}{\pi^2 c^3 \hbar^3} \left( \frac{\pi^4}{15} + 120 \zeta(5) \alpha_0 k_B T \right), \\
		P_{(III)} &= \frac{(k_B T)^4}{3\pi^2 c^3 \hbar^3} \left( \frac{\pi^4}{15} + 72\zeta(5)\alpha_0 k_B T \right).
	\end{align}
	
	These expressions define an additional  quantity $\rho_{X}(T)$, which gives a supplementary contribution to the total radiation energy density during BBN. This deviation is responsible for a modification in the expansion rate, which affects the freeze-out temperature $T_f$ of weak particle decoupling. A perturbation in $\rho$ also leads to a shift in the Hubble parameter, which can be linked to the freeze-out deviation temperature according to \cite{Capozziello}
	\begin{equation}
		\frac{\delta H}{H} = \frac{1}{2} \frac{\delta \rho}{\rho} \Rightarrow \frac{\delta T_f}{T_f} = \frac{1}{5} \frac{\delta H}{H}.
	\end{equation}
	BBN constraints on helium abundance impose an upper limit on the freeze-out temperature shift, leading to an upper strong constraint on the energy density correction
	\begin{equation}
		\left| \frac{\delta T_f}{T_f} \right| < 4.7 \times 10^{-4} \Rightarrow  \rho_{X} < 1.35 \times 10^{-3}\;\mathrm{MeV}^4.
	\end{equation}
	
	\subsection{Numerical integration with \texttt{PRyMordial}}
	\label{prym_workflow}
	
	We can incorporate the additional contribution to the standard radiation fluid dynamics coming from the non-commutative models into the BBN theory through the  \href{https://github.com/vallima/PRyMordial/tree/main}{\texttt{PRyMordial}} library. This python code simulates the nucleosynthesis process by directly integrating the modified thermodynamic quantities into the calculations of primordial nuclei abundances \cite{primordial}. These are estimated by numerically solving the Boltzmann equations for the nuclear species, evaluating the temperature, scale factor, and baryon number density evolution
	\begin{equation}
		\left(\frac{dT}{dt},  \frac{da}{dt},  \frac{dn_b}{dt} = -3Hn_b\right),
	\end{equation}
	throughout the BBN epoch. The program integrates both Standard Model and New Physics scenarios,  evaluating the coupled radiation and matter content from $T_\gamma = 10~\mathrm{MeV}$ to the keV scale, accounting for incomplete neutrino decoupling which further impacts the neutrino temperature evolution $T_\nu$. The program workflow is outlined in what follows.

	\paragraph{Directory layout.}
	The public repository is organised in two main folders:
	\texttt{PRyM/}, which contains the solver, thermodynamics and weak–interaction
	kernels, and \texttt{PRyMrates/}, which includes tabulated NACRE II  \cite{nacre2}
	thermonuclear reaction rates, together with their covariance matrices.
	
	\paragraph{Physical setup and ODE system.}
	At run-time the code builds the state vector
	\[
	\mathbf{y}(t)=
	\bigl(
	X_{n}(t),\,X_{p}(t),\ldots,X_{{}^{7}\mathrm{Li}}(t);\;
	T_\gamma(t),\,T_\nu(t);\;
	n_b(t),\;a(t)
	\bigr),
	\]
	where by $X_i$ we have denoted the nuclear mass fractions, $T_\gamma$ and $T_\nu$ are the
	photon and three neutrino temperature flavours, $n_b$ is the baryon
	number density, and $a$ is the FLRW scale factor.  The evolution is
	governed by the coupled system of equations
	\begin{align}
		\frac{dX_i}{dt} &=
		\sum_{j,k} N^{(i)}_{jk}\,\lambda_{jk}(T_\gamma)\,
		X_j X_k
		- X_i\!\sum_{\ell}\lambda_{i\ell}(T_\gamma)X_\ell,
		\label{eq:network}\\[4pt]
		\frac{dT_\gamma}{dt} &=
		-\frac{T_\gamma}{a}\frac{da}{dt}
		-\frac{\,\dot q_{\rm e^\pm}(T_\gamma)\,}{3
			\bigl(\rho_\gamma+P_\gamma\bigr)} ,
		\\
		\frac{dT_\nu}{dt} &=
		-H\,T_\nu + \mathcal{C}_\nu(T_\gamma,T_\nu),
		\label{eq:temps}\\[4pt]
		\frac{da}{dt} &= a\,H(T_\gamma,T_\nu,X_i),
		\\
		\frac{dn_b}{dt} &= -3H\,n_b,
		\label{eq:cosmo}
	\end{align}
	where $N^{(i)}_{jk}$ denotes the nuclear stoichiometry,
	$\lambda_{jk}$ are temperature-dependent reaction rates, read from
	\texttt{PRyMrates/}, $H=\sqrt{8\pi G\rho_{\rm tot}/3}$ is the Hubble
	parameter, $\dot q_{e^\pm}$ accounts for $e^\pm$ annihilation heating,
	and $\mathcal{C}_\nu$ implements finite-temperature QED and incomplete
	neutrino decoupling corrections.
	
	\paragraph{Integrator.}
	The stiff system (\ref{eq:network})–(\ref{eq:cosmo}) is solved from
	$T_\gamma\simeq10\ \mathrm{MeV}$ down to $T_\gamma\simeq1\ \mathrm{keV}$
	with \texttt{scipy.integrate.solve\_ivp} and \texttt{LSODA} and \texttt{Radau} methods,
	automatically adapting the step size and dense output to resolve the
	faster nuclear reactions, as well as the slower cosmological cooling
	background.
	
	The general method used in this investigation is described in the following workflow summary:
	\begin{enumerate}
		\item \emph{Initialisation:} constants, rate interpolants, and user-selected flags (12 vs.\ 63 reactions, NP sector, incomplete decoupling effects in $a(T)$) are
		handled by \texttt{PRyM\_init.py}.
		\item \emph{Integration:} Eqs.~(\ref{eq:network})–(\ref{eq:cosmo})
		with \texttt{solve\_ivp}, storing intermediate time-steps for
		diagnostics.
		\item \emph{Freeze-out detection:} once all forward \(p\leftrightarrow n\)
		and nuclear rates satisfy
		$\Gamma/H<10^{-3}$, abundances are considered frozen.
		\item\emph{Post-processing:}
		The final state is evaluated by \texttt{PRyM\_results}, which returns the observables
		\begin{align}
			\bigl\{
			& N_{\mathrm{eff}},\;
			\Omega_\nu h^2 \times 10^{-6},\;
			\sum m_\nu/\Omega_\nu h^2,\;
			Y_p^{\text{(CMB)}}, \; \notag \\
			& Y_p^{\text{(BBN)}},\;
			\mathrm{D/H} \times 10^5,\;
			{}^3\mathrm{He/H}\times 10^5,\;
			{}^7\mathrm{Li/H} \times 10^{10}
			\bigr\},
		\end{align}
		written to a plain-text table and returned for further parameter or likelihood analyses.
	\end{enumerate}
	
	The non-commutative effects are included through the \texttt{New-Physics}
	interface in \texttt{PRyM\_thermo.py}, which modifies the Hubble expansion rate as $ H(T) \propto \sqrt{\rho_{\mathrm{tot}}} $, with
	\begin{equation}
		\rho_{\mathrm{tot}}(T) = \rho_\gamma(T) + \rho_e(T) + 3\rho_\nu(T) + \rho_{X}(T).
	\end{equation}
	
	The effects of considering a deformed photon gas into the early time evolution of the Universe induces a shift in the neutron freeze-out temperature, thus modifying the neutron-to-proton ratio and the nuclei abundances. The updated predictions for $Y_p$ and D/H can be used to further constrain the deformation parameters $\lambda$, $\beta_0$, and $\alpha_0$ by comparing them to the standard predictions.	
			
\section{Methodology, statistical analysis, and BBN constraints on noncommutativity effects} \label{Stat}
	
	In the context of non-standard BBN models, the posterior distribution often shows non-Gaussian behaviour. Monte Carlo methods provide a numerical solution by approximating expectations through random sampling. This Section presents the main methods used in the estimation of the deformation parameters and the model selection criteria. 
	
	\subsection{MCMC parameter estimation} \label{mcmc}
	
	To generate samples from the posterior, we employ Markov Chain Monte Carlo (MCMC) methods, which construct a chain of parameter values whose distribution converges to the estimated parameter's posterior. Specifically, we use the Metropolis–Hastings algorithm, which is suitable for simulating the nuclear synthesis process, for which the likelihood evaluation takes a long time as it involves numerically solving a stiff system of coupled differential equations.
	
	In order to ensure the convergence of the chains, we used a chi-squared statistic defined by
	\begin{equation}
		\chi^2 = \sum_i \frac{(A_{i,\mathrm{obs}} - A_{i,\mathrm{th}})^2}{\sigma_i^2},
		\label{chi}
	\end{equation}
	where  $A_{i,obs} $ refers to the observed and $A_{i,th}$ to the theoretical primordial abundances of helium-4, and deuterium nuclei. The observational data used are \cite{pdg}
	\begin{eqnarray}
		&&Y_p^{\text{(CMB)}} = 0.245 \pm 0.003, \\
		&&\mathrm{D/H} = (2.547 \pm 0.029) \times 10^{-5}. 
		%^3\mathrm{He}/\mathrm{H} &\approx& (1.08 \pm 0.12) \times 10^{-5}\qquad \cite{pdg}.
	\end{eqnarray}
	The values $A_{i,\mathrm{obs}}$ correspond to abundances predicted by the inclusion of a deformed dispersion relation for the photon gas. The uncertainties $\sigma_i$ reflect the variance of the data.

	We incorporate the modified dispersion relations into the \href{https://github.com/teomatei22/prynce}{\texttt{PRyNCe}} framework, an extension of the \texttt{PRyMordial} BBN code. Each dispersion model modifies the radiation energy density and pressure, altering the predicted primordial abundances of light elements. Since \texttt{PRyMordial} uses natural units as a convention, the temperature is measured in MeV and the energy density in MeV$^4$, so the deformation parameters $\lambda$ and $\alpha_0$ are measured in MeV$^{-1}$ and $\beta_0$ in MeV$^{-2}$, with $k_B = 1$. We constrain the deformation parameters by using an affine-invariant MCMC sampler called \href{https://github.com/dfm/emcee}{\texttt{emcee}}, by running 8 synchronized walkers for 3000 steps each.
	
	The following uniform priors were assigned for each dispersion parameter,
	\begin{eqnarray}
		\lambda &\in& [10^{-10},9\times10^{-3}] \; \rm{MeV}^{-1}, \notag \\
		\beta_0 &\in& [10^{-9},\;1 \times 10^{-3}] \; \rm{MeV}^{-2}, \notag \\
		\alpha_0 &\in& [10^{-10},2 \times 10^{-4}]  \; \rm{MeV}^{-1}, \notag
	\end{eqnarray}
	which were selected based on their ability to avoid instabilities in the numerical evaluation of \texttt{PRyMordial}, while estimating physical results for the nucleosynthesis process.
	
	Two additional constraints are imposed to ensure the deformed parameter validity in the oncoming synthesis process. First, we require that the total energy density at freeze-out must remain below 
	\begin{equation}
		\rho_{\text{deform}}(T_F) < 0.00135  \; \rm{MeV}^{4}.
	\end{equation} 
	
	Second, the equation of state parameter for the modified radiation sector lies within physical bounds at freeze-out temperature $T_F = 0.5\,\text{MeV}$ and below,
	\begin{equation}
		-1 < w_F = \frac{p_F}{\rho_F} < 1.
	\end{equation}
	Any parameter values violating these conditions are assigned zero likelihood, which corresponds to a log-prior value of $-\infty$.
	
	Walkers are initialized uniformly across the prior space, and the sampling is parallelized using the multiprocessing backend of \texttt{PRyNCe} to accelerate evaluation, as each likelihood computation involves a full BBN calculation. The first 10 percent of samples, corresponding to 300 steps per walker, are discarded as burn-in. The MCMC analysis incorporates uncertainties in 12 key nuclear reaction rates, neutron lifetime is given as $\tau_n = 879.4 \pm 0.2$ s, and baryon density as $\Omega_b h^2 = 0.0230 \pm 0.0001$, respectively.

	\subsection{Convergence Analysis}

    One of the most used methods for determining the convergence of the MCMC process is the Gelman-Rubin R statistic, whose basic idea is to run multiple 
	independent chains in parallel and compare the variability 
	within each chain to the variability between chains. 
	If all chains have reached the same stationary distribution, 
	these two variabilities should be similar.  
	
	The within-chain variance is defined as
	\begin{equation}
		W = \frac{1}{M} \sum_{m=1}^{M} s_m^2, 
	\end{equation}
	where $ M  $ is the number of parallel chains and  $ s_m^2  $ is the sample variance of chain  $ m  $.  
	
	The between-chain variance is
	\begin{equation}
	B = \frac{N}{M-1} \sum_{m=1}^{M} \left(\bar{\theta}_m - \bar{\theta}\right)^2,
	\end{equation}
	where  $ N  $ is the length of the chains, $\theta $ is the sampling a parameter, $ \bar{\theta}_m  $ is the mean of chain  $ m  $, and 
	 $ \bar{\theta} = \tfrac{1}{M} \sum_{m=1}^M \bar{\theta}_m  $ 
	is the overall mean across all chains.
	
	An unbiased estimate of the marginal posterior variance of  $\theta$ can be obtained by combining the within-chain variance $W$ and between-chain variance $B$,
	\begin{equation}
	\hat{V} = \frac{N-1}{N} W + \frac{1}{N} B.
	\end{equation}
	This quantity accounts for both short-term variability within individual chains and long-term variability across chains. When the chains explore different regions in the parameter space, 
	the between-chain term $B$ becomes large, so that $\hat{V}$ exceeds $W$. Conversely, if the chains explore the same stationary distribution, the contribution of $B$ remains small and $\hat{V}$ approaches $W$.
	
	The variance estimate is used to construct the Gelman--Rubin statistic, defined as \cite{Gelman}
	\begin{equation}
	\hat{R} = \sqrt{\frac{\hat{V}}{W}}.
	\end{equation}
	Intuitively, $\hat{R}$ quantifies how dispersed the posterior variance of the estimated parameter would become if the chains were extended indefinitely. An $\hat{R}$ close to one suggests that it is unlikely to obtain a different variance estimate by further sampling from the parameter distribution. In practice, values of  $ \hat{R} < 1.01  $ are a common indicator for reaching
	convergence \cite{vehtari}. 
	
	We applied the Gelman-Rubin statistic to our three non-commutativity models and obtained the following $\hat{R}$ values, which satisfy the convergence criterion,
	\begin{eqnarray}
		\text{Model I:} \;& \hat{R} = 1.008, \nonumber \\
		\text{Model II:} \;& \hat{R} = 1.002, \nonumber \\
		\text{Model III:}\; & \hat{R} = 1.014. \nonumber
	\end{eqnarray}
	
\begin{figure}[htbp!]
		\centering
		\includegraphics[scale = 0.285]{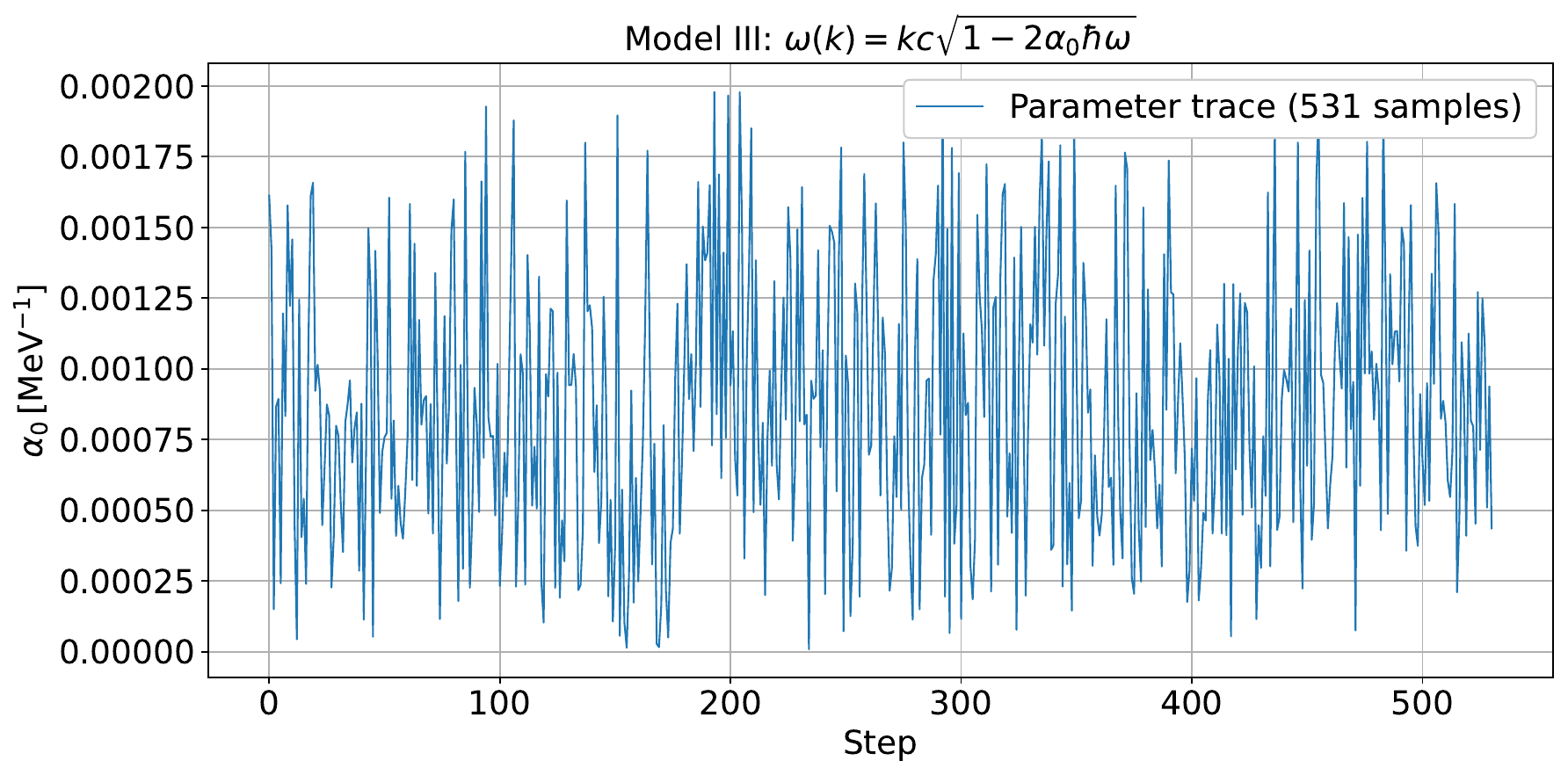}
		\caption{Trace plot showing chain convergence for $\alpha_0$ dispersion parameter.}
		\label{trace}
	\end{figure}

\begin{figure}[htbp!]
		\centering
		\includegraphics[scale = 0.35]{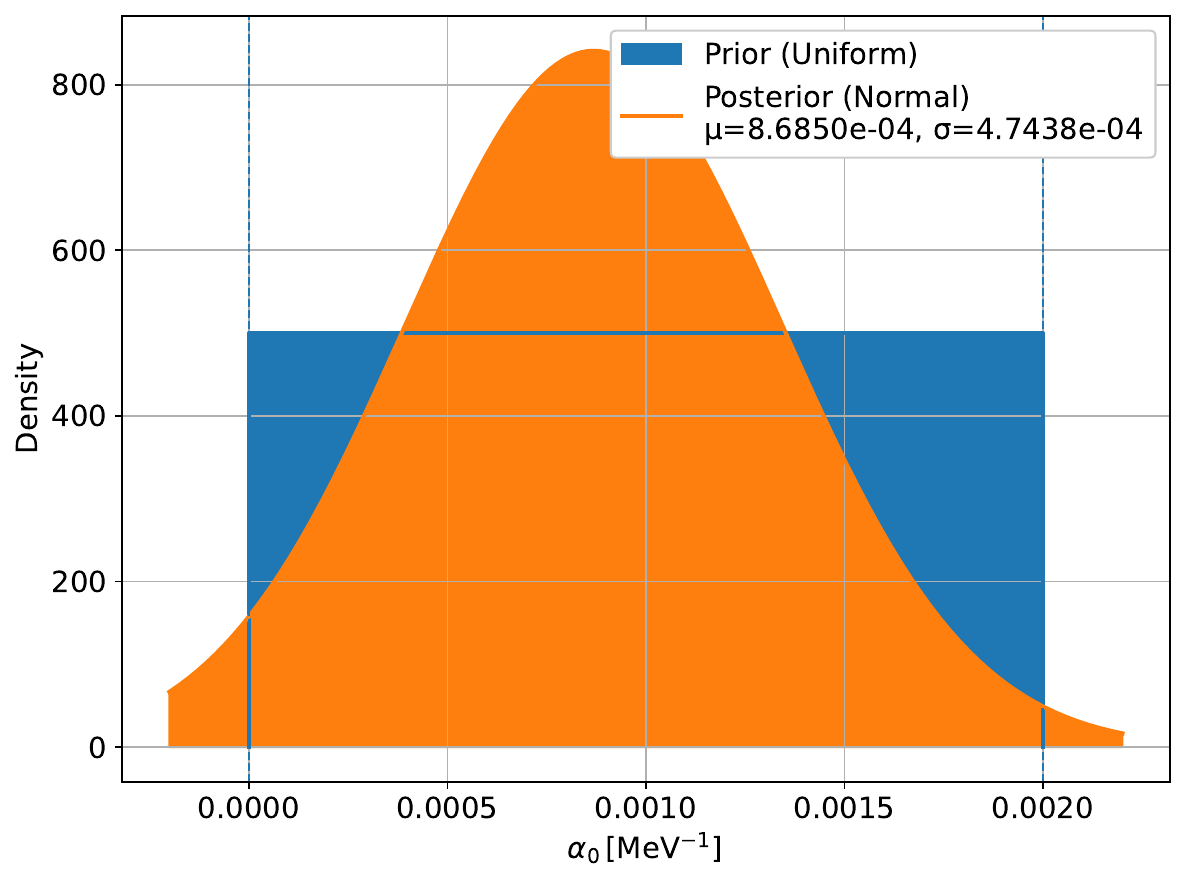}
		\caption{Prior and posterior distribution overlay for the $\alpha_0$ parameter in Model III.}
		\label{prior_posterior}
	\end{figure}

	It is important to note that  $ \hat{R}  $ is not a sufficient diagnostic on its own for assessing convergence.
	Visual inspection of trace plots and the examination of posterior convergence with respect to the chosen prior are necessary for verifying that the parameter space was efficiently explored. In this regard, Fig.~\ref{trace} shows a representative trace plot for Model III, which has an $\hat{R}$ value exceeding the threshold, but whose chains nevertheless converge, while Fig.~\ref{prior_posterior} displays its prior and posterior distribution overlay.  
	
	As one can see from the prior-posterior overlaying distributions, by sampling dispersion parameters that satisfy the physical conditions imposed by nucleosynthesis, the chains converge to a stationary distribution regardless of the chosen uniform priors. This behaviour is consistent across all three models, confirming that each chain converges successfully to its respective posterior distribution.
	 
	\subsection{Statistical analysis} \label{stat}
	 
	In many fitting procedures, the reduced chi-squared statistic is used as a measure of goodness-of-fit,
	\begin{equation}
		\chi^2_{\nu} = \frac{\chi^2}{\nu},
	\end{equation}
	 where $\nu = n_{data} - k$ is the number of degrees of freedom, with $n_{data}$ the number of observational data points and $k$ the number of fitted parameters. A value of $\chi^2_{red} \approx 1$ typically indicates that the model describes the data within the expected statistical noise.
	
	However, in cases when the number of degrees of freedom is small, the $\chi^2_{\nu}$ metric becomes unreliable as the chi-squared distribution for small $\nu$ is highly sensitive to noise. As a result, deviations from the expected value of 1 can arise from statistical fluctuations.
	
	To properly interpret the fit in such situations we use the chi-squared probability,
	\begin{equation}
		P(\chi^2_{obs}; \nu) = \int_{\chi^2_{obs}}^{\infty} p_\chi(\chi^2; \nu) \, d(\chi^2),
	\end{equation}
	where $p_\chi(\chi^2; \nu)$ is the chi-squared probability density function
	\begin{equation}
		p_\chi(\chi^2; \nu) = \frac{1}{2^{\nu/2} \Gamma(\nu/2)} (\chi^2)^{(\nu - 2)/2} e^{-\chi^2/2}.
	\end{equation}
	This quantity gives the likelihood of obtaining a value of $\chi^2$ that exceeds the observed $\chi^2$ associated to a model parameter \cite{stat}. Values of $P$ near 0.5 indicate that there is no significant underfitting or overfitting of the data.
	
	Table~\ref{pvalue} displays the $\chi^2$ values corresponding to several representative p-values for different degrees of freedom. It is worth noting that our analysis is constructed in a low degree of freedom regime, with $\nu = 1$ as we consider only two primordial abundances as number of data points and one model parameter corresponding to the dispersion coefficients of the non-commutativity models. In this regime, the reduced chi-squared values as low as 0.455 still correspond to statistically acceptable fits ($P = 50\%$), unlike the high degree of freedom case where  $\chi^2_{\nu} \approx 1$ is expected.
	
		\begin{table}[h!]
		\centering
		\begin{tabular}{|c|c|c|c|c|c|}
			%	\toprule
			\hline
			
			$\nu$ & $P=0.70$ & $P=0.60$ & $P=0.50$ & $P=0.40$ & $P=0.30$ \\
			%	\midrule
			\hline
			1  & 0.148 & 0.275 & 0.455 & 1.074 & 1.642 \\
			\hline
			2  & 0.357 & 0.511 & 0.693 & 1.204 & 1.609 \\
			\hline
			3  & 0.475 & 0.623 & 0.789 & 1.222 & 1.547 \\
			\hline
			\vdots & \vdots & \vdots & \vdots & \vdots & \vdots \\
			\hline
			20 & 0.813 & 0.890 & 0.967 & 1.048 & 1.139 \\
			\hline
			%	\bottomrule
			
		\end{tabular}
		\caption{Observed $\chi^2$ values corresponding to probability values for various degrees of freedom $\nu$, adapted from \cite{stat} (p. 256-257).}
		\label{pvalue}
	\end{table}

	We restrict our analysis to this minimal configuration because the model parameters are highly sensitive to observational uncertainties. Therefore, as  helium-3 and lithium abundances have large observational uncertainties, their inclusion in the $\chi^2$ function would make it difficult to detect the small dispersion effects we are investigating.

 	In addition to the p-value,  we employed the Akaike Information Criterion \cite{aic1} and the Bayesian Information Criterion \cite{bic1, bic2} to compare the different deformed photon gas models. Both information criteria evaluate model performance by examining the goodness-of-fit and model complexity, providing an important additional metric for the statistical analysis.
 	
 	\begin{figure*}[htbp!]
 		\centering
 		\includegraphics[scale = 0.335]{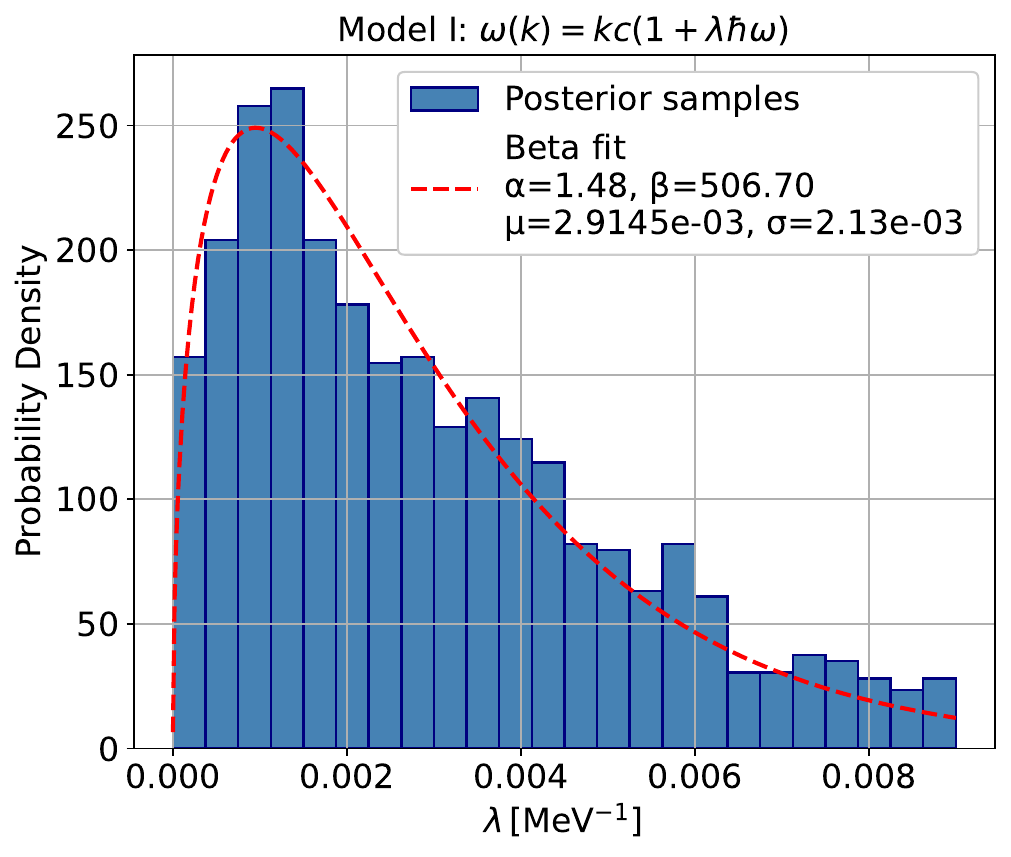}
 		\includegraphics[scale = 0.335]{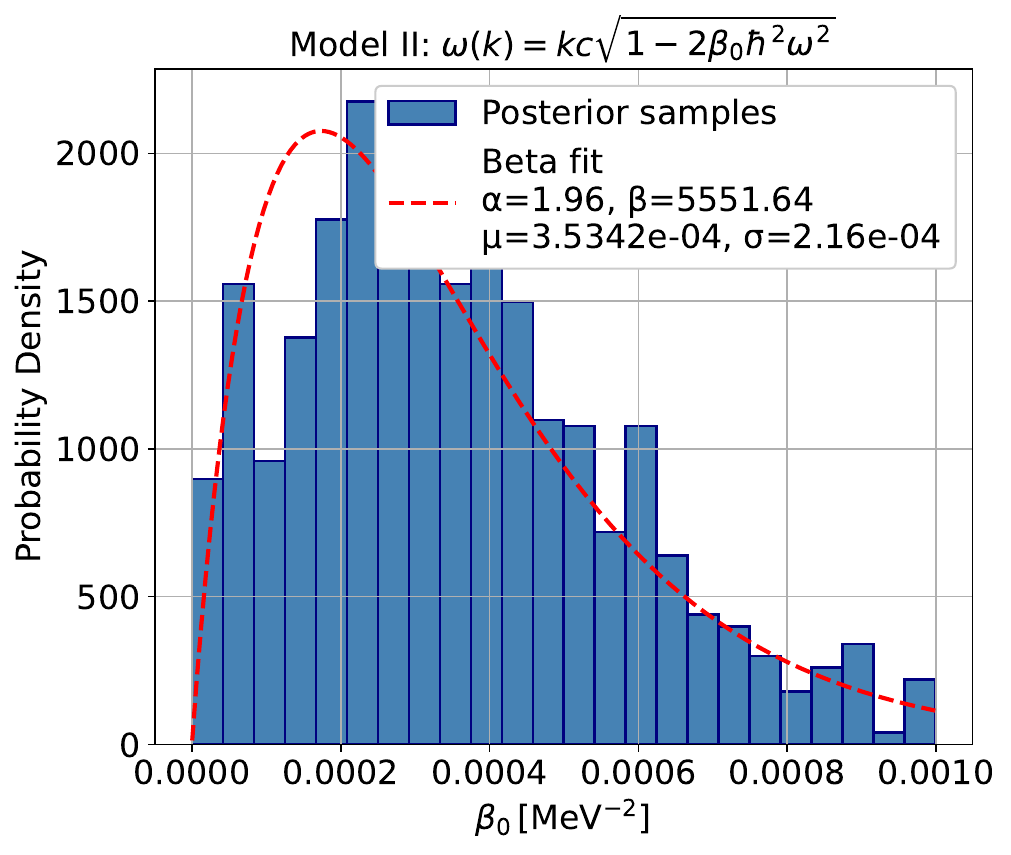}
 		\includegraphics[scale = 0.335]{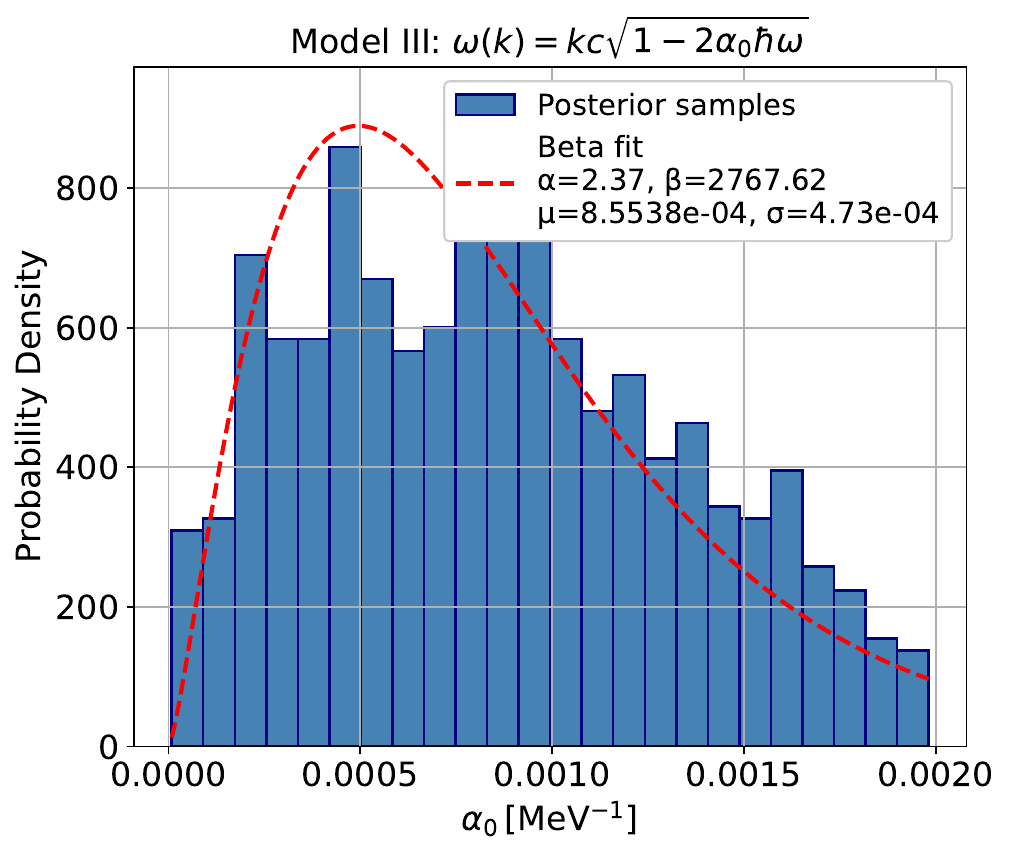}
 		\caption{The posterior distributions or the deformed photon gas parameter models $\theta = \{\lambda, \beta_0, \alpha_0\}$, estimated with the MCMC algorithm. The plot density $\rho(\theta)$ is visualized against the deformed parameters $\theta$. }
 		\label{posteriors}
 	\end{figure*}
 	
 	\begin{table*}[htbp!]
 	\centering
 	\renewcommand{\arraystretch}{1.6}
 	\setlength{\tabcolsep}{4pt}
 	\begin{tabular}{|l|c|c|c|c|c|c|}
 		%		\toprule
 		\hline
 		Dispersion Parameter & $Y_p^{\text{CMB}}$ & D/H $\times 10^5$ & $^3$He/H $\times 10^5$ & $^7$Li/H $\times 10^{10}$ & $\rm{N_{eff}}$ & $\Omega_\nu h^2 \times 10^6$ \\
 		%\midrule
 		\hline
 		$\lambda \;\;= (2.91 \pm 2.13) \times 10^{-3} \; [\rm{MeV}^{-1}]$ & $0.2494^{+0.0021}_{-0.0021}$ & $2.5737^{+0.6646}_{-0.6646}$  & $1.0544^{+0.1298}_{-0.1298}$ & $5.8072^{+2.1291}_{-2.1291}$ & $3.1743^{+0.0984}_{-0.0984}$ & $5.9428^{+0.1843}_{-0.1843}$ \\
 		\hline
 		$\beta_0 = (3.53 \pm 2.16) \times 10^{-4} \; [\rm{MeV}^{-2}]$ & $0.2425^{+0.0023}_{-0.0023}$ & $2.3535^{+0.3983}_{-0.3983}$ & $1.0243^{+0.1074}_{-0.1074}$ & $5.9589^{+1.5432}_{-1.5432}$ & $2.8586^{+0.1084}_{-0.1084}$ & $5.3515^{+0.2030}_{-0.2030}$ \\
 		\hline
 		$\alpha_0 = (8.55 \pm 4.73) \times 10^{-4} \; [\rm{MeV}^{-1}]$ & $0.2483^{+0.0019}_{-0.0019}$ & $2.2036^{+0.2206}_{-0.2206}$ & $1.0065^{+0.0989}_{-0.0989}$ & $6.4851^{+1.2790}_{-1.2790}$ & $3.2076^{+0.0926}_{-0.0926}$ & $6.0050^{+0.1733}_{-0.1733}$ \\
 		\hline
 	\end{tabular}
 	\caption{Results from the nucleosynthesis computation with \texttt{PRyMordial} for the three cases of dispersion for the deformed photon gas. }
 	\label{table_results}
 \end{table*}

	The AIC is defined as
	\begin{equation}
		\mathrm{AIC} = 2k - 2\ln\mathcal{L},
	\end{equation}
	where  $k $ is the number of model parameters and  $\mathcal{L} $ the maximum likelihood. Models with lower AIC values are favoured as they minimize information loss.
	
	Similarly, the BIC is given by
	\begin{equation}
		\mathrm{BIC} = k \ln n - 2\ln\mathcal{L},
	\end{equation}
	where  $n $ represents the number of data points. Compared to AIC, the BIC is  stricter for large datasets.
	
	Model performance is determined by the differences of the AIC and BIC coefficients with respect to the best-performing model,
	\begin{eqnarray}
		\Delta_{\mathrm{AIC}} &=& \mathrm{AIC}_i - \mathrm{AIC}_{\mathrm{min}}, \\
		\Delta_{\mathrm{BIC}} &=& \mathrm{BIC}_i - \mathrm{BIC}_{\mathrm{min}}.
	\end{eqnarray}

	Following the standard interpretation in \cite{aic2}, models with  $\Delta_{\mathrm{AIC}} \leq 2 $ and  $\Delta_{\mathrm{BIC}} \leq 2 $ are considered to have substantial support. Altogether, models with  $4 \leq \Delta_{\mathrm{AIC}} \leq 7 $ or  $2 \leq \Delta_{\mathrm{BIC}} \leq 6 $ are disfavoured by the data, while models with  $\Delta_{\mathrm{AIC}} > 10 $ and  $\Delta_{\mathrm{BIC}} > 6 $ indicate strong evidence against the data.
	
\subsection{Results of the PRyNce framework}
	
	The methodology described in subsection \ref{mcmc} was used to obtain the results shown in Fig.~\ref{posteriors} for the posterior distributions of the deformation parameters. The parameter estimates show Beta distributions for all three dispersion models, indicating an expected preference for small values. Specifically, the first dispersion model is described by a Beta distribution with shape parameters $ \alpha = 1.4808 $ and $ \beta = 506.70 $, while the second model follows a Beta distribution with $ \alpha = 1.9612 $ and $ \beta = 5551.64 $ and finally, the third model distribution has the parameters $ \alpha = 2.3693 $ and $ \beta = 2767.62 $, respectively. 
	
	For a Beta distribution characterized by $\alpha$ and $\beta$ parameters, the mean and standard deviation are given by
	\begin{equation}
	\mu = \frac{\alpha}{\alpha + \beta}, \quad
	\sigma = \sqrt{ \frac{\alpha \beta}{(\alpha + \beta)^2 (\alpha + \beta + 1)} }.
	\end{equation}
	We find after applying these expressions $\mu_{\lambda} = 2.9145 \times 10^{-3} $ and $\sigma_{\lambda} = 2.13 \times 10^{-3}$ for the first model, $\mu_{\beta_0} = 3.5342 \times 10^{-4} $ and $\sigma_{\beta_0} = 2.16 \times 10^{-4}$ for the second model and  $\mu_{\alpha_0} = 8.5538 \times 10^{-4}$ and $\sigma_{\alpha_0} = 4.73 \times 10^{-4}$ for the third model, respectively. These results together with the predictions of Helium-4 and deuterium abundances, which were directly used in the evaluation of the chi-squared function, and are displayed in Table~\ref{table_results}, along with other quantities computed by \texttt{PRyMordial}, such as the Helium-3 and Lithium-7 yields. Moreover, the effective number of degrees of freedom for neutrino species $N_{\text{eff}}$ is obtained together with the neutrino density parameter, described by the following relation in the late-time regime
	\begin{equation}
		\Omega_\nu h^2 = \frac{\rho_\nu}{\rho_\text{crit}/h^2}.
	\end{equation}
	
	\begin{figure*}[htbp!]
		\centering
		\includegraphics[scale = 0.4]{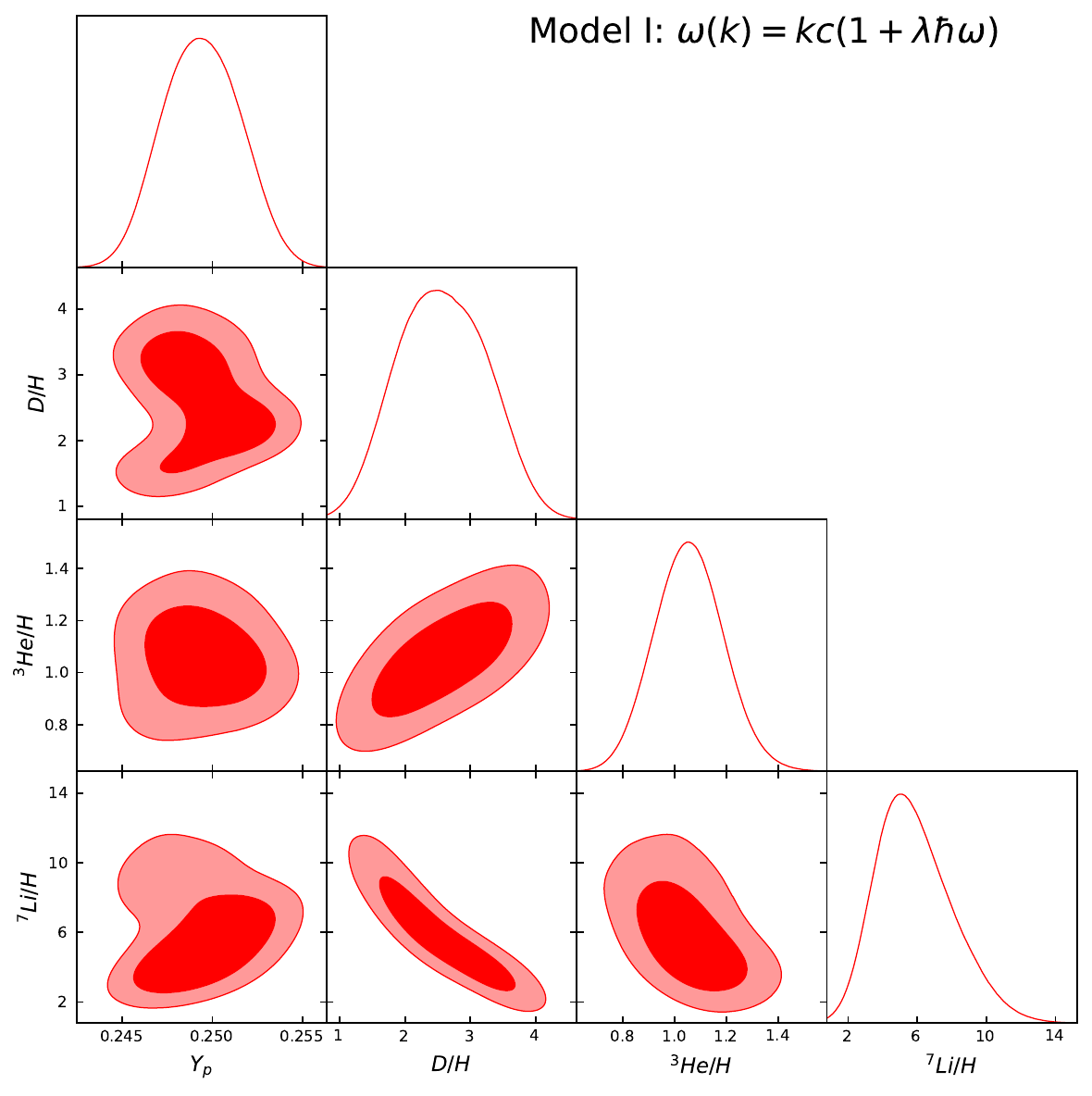}
		\includegraphics[scale = 0.4]{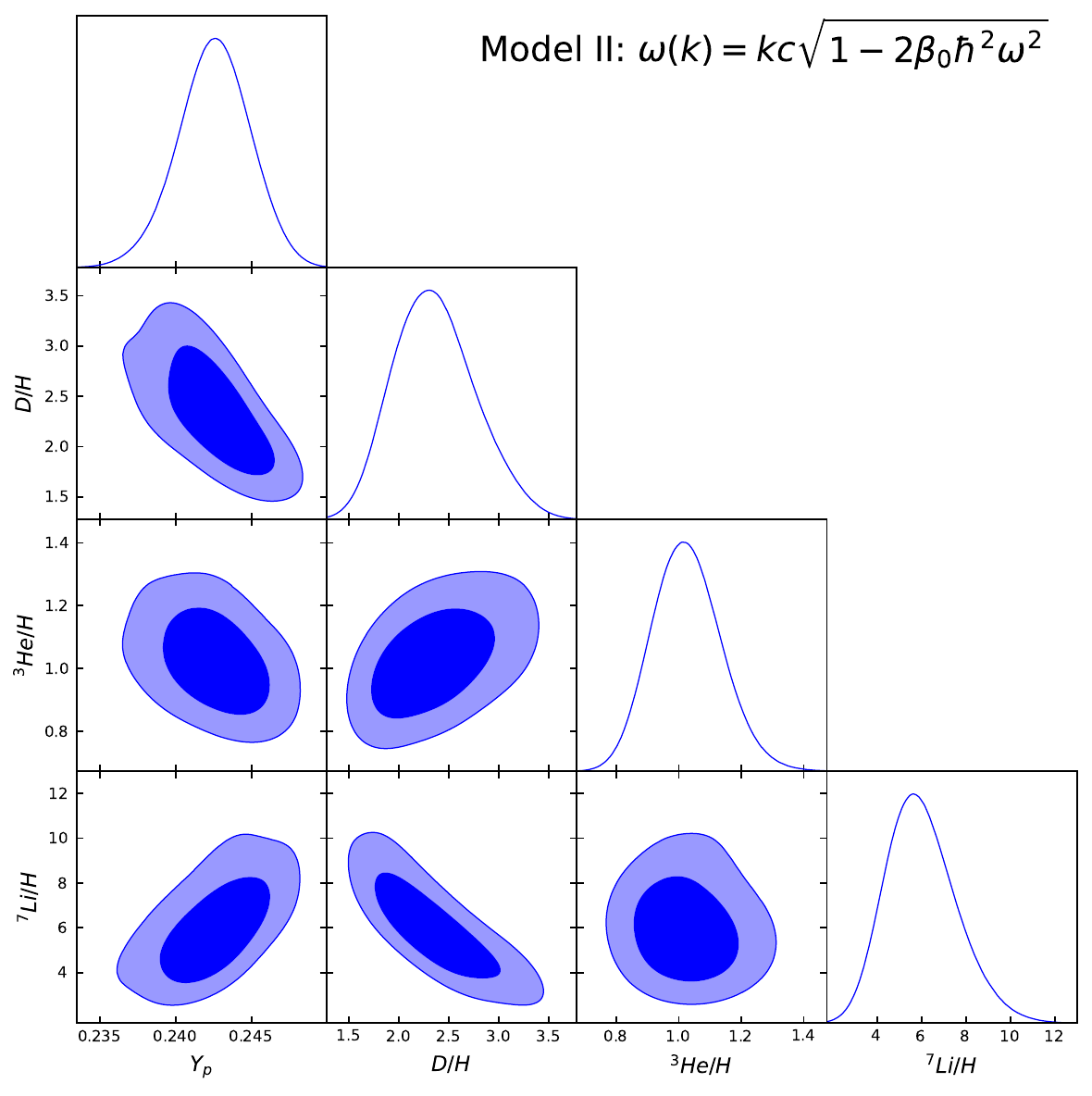}
		\includegraphics[scale = 0.4]{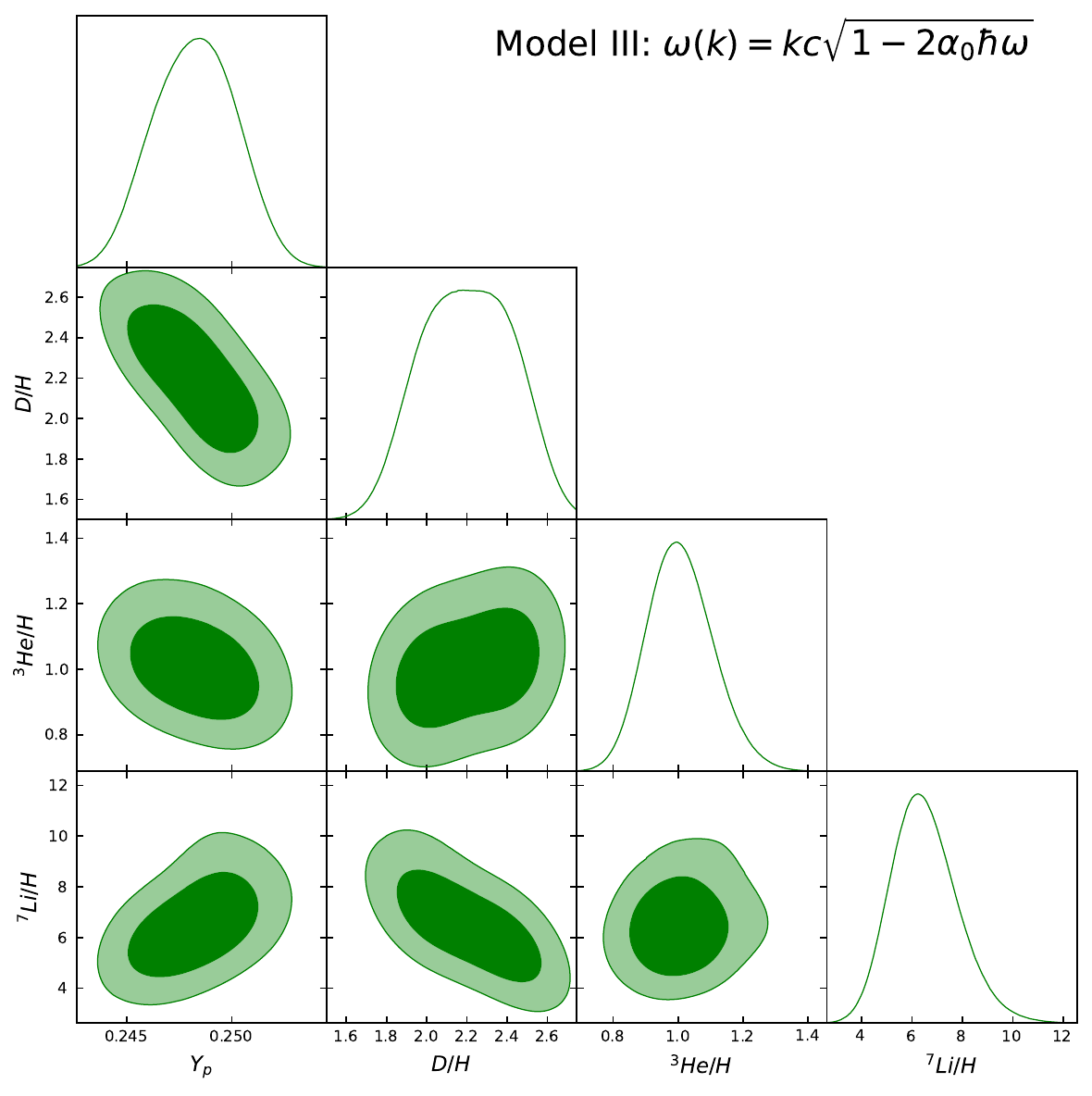}
		\caption{The resulted nuclei abundances considering the three models of deformed photon gas (first model - upper left, second model - upper right, third model - bottom). The model parameters have the mean and standard deviations from Table.~\ref{table_results}.}
		\label{corners}
	\end{figure*}	
	
	The standard value $ N_{\text{eff}}^{\text{SM}}= 3.046 $ highlights the presence of three active neutrino species, with a small correction from non-instantaneous decoupling. Our analysis finds $ N_{\text{eff}} = 3.174 \pm 0.098$ for the first model, $ 2.859 \pm 0.108 $ for the second model, and $ 3.208 \pm 0.093 $ for the third model. 
	
	To interpret these results, we define $\Delta N_\nu = N_{\text{eff}} - N_{\text{eff}}^{\rm SM}$, resulting in $\Delta N_{\nu(I)} = 0.128 \pm 0.098$, $\Delta N_{\nu(II)} = -0.187 \pm 0.108$ and $\Delta N_{\nu(III)} = 0.162 \pm 0.093$, respectively. The negative corrections from the standard Stefan-Boltzmann law in Model I and the positive corrections in Model III reflect an increase in the relativistic energy density during BBN, while the positive corrections in Model II indicate a suppressed relativistic sector. 
	All results remain consistent with Planck observations, which allows for deviations up to $ \Delta N_\nu \lesssim 0.3$ at the 95\% confidence level.
	
	Fig.~\ref{corners} presents the posterior distribution of the nuclei abundances, showing expected correlations between the primordial element formations. The corner plots were obtained using \href{https://getdist.readthedocs.io}{\texttt{getdist}} with \texttt{smooth\_scale\_1D = 0.50} and \texttt{smooth\_scale\_2D = 0.75}. 
	
	\begin{table}[htbp!]
		\centering
		\begin{tabular}{|l|c|c|c|c|}
			\hline
			Parameter & $Y_p^\text{CMB}$ & $\text{D/H}$ & $^3\text{He/H}$ & $^7\text{Li/H}$ \\
			\hline			%\midrule
			\multicolumn{5}{l}{Model I ($\lambda$)} \\
			\hline
			$Y_p^{\text{CMB}}$ & 1.000 & -0.187 & -0.085 & 0.109 \\
			$\text{D/H}$ & -0.187 & 1.000 & 0.690 & -0.898 \\
			$^3\text{He/H}$ & -0.085 & 0.690 & 1.000 & -0.466 \\
			$^7\text{Li/H}$ & 0.109 & -0.898 & -0.466 & 1.000 \\
			\hline			%\midrule
			\multicolumn{5}{l}{Model II ($\beta_0$)} \\
			\hline
			$Y_p^\text{CMB}$ & 1.000 & -0.672 & -0.370 & 0.576 \\
			$\text{D/H}$ & -0.672 & 1.000 & 0.473 & -0.802 \\
			$^3\text{He/H}$ & -0.370 & 0.473 & 1.000 & -0.089 \\
			$^7\text{Li/H}$ & 0.576 & -0.802 & -0.089 & 1.000 \\
			\midrule
			\multicolumn{5}{l}{Model III ($\alpha_0$)} \\
			\hline
			$Y_p^\text{CMB}$ & 1.000 & -0.376 & -0.373 & 0.465 \\
			$\text{D/H}$ & -0.376 & 1.000 & 0.825 & -0.700 \\
			$^3\text{He/H}$ & -0.373 & 0.825 & 1.000 & -0.628 \\
			$^7\text{Li/H}$ & 0.465 & -0.700 & -0.628 & 1.000 \\
			\bottomrule
		\end{tabular}
		\caption{Correlation coefficients between abundance parameters for the three dispersion models.}
		\label{table_correlations}
	\end{table}
	
	The interdependencies between the primordial abundances can be expressed by calculating the 
	correlation coefficient $r_{ij}$. This quantity describes the linear relationship between two parameters and is given by
	\begin{equation}
		r_{ij} = \frac{\text{Cov}(i, j)}{\sigma_i \sigma_j},
	\end{equation}
	where $\text{Cov}(i, j)$ is the covariance between the $i$ and $j$ parameters and $\sigma_i$ and $\sigma_j$ are their respective standard deviations. Values near $\pm 1$ indicate strong linear correlation, while values near 0 imply statistical independence.
	
	The correlation coefficients are displayed in Table~\ref{table_correlations}, and show similar values across all three dispersion models. This suggests that the underlying nuclear physics dominates over the specific form of the dispersion relation, especially at low temperatures, when the nucleosynthesis occurs.
	All models show a moderate negative correlation between $ Y_p^{\text{CMB}}$ and D/H, ranging from $ r = -0.187 $ (Model I) to $r = -0.674$ (Model II). A strong negative correlation is consistently observed between $^7\text{Li}/\text{H}$ and D/H across all models, with values of $r = -0.898, -0.802$ and $-0.700$ for Models I, II, and III, respectively. Additionally, a strong positive correlation exists between D/H and $^3\text{He}/\text{H} $, with correlation coefficients of $r = 0.690, 0.473$ and $0.825$ for the three models.

	The statistical analysis described in Section \ref{stat} provides the model comparison  coefficients presented in Table~\ref{coeff} for the three spacetime noncommutativity models. The quantities  $\Delta $AIC and  $\Delta $BIC represent differences relative to the best-performing model, which is identified by having the lowest AIC and BIC values.
	
	\begin{table}[htbp!]
		\vspace{0.5cm}
		\centering
			\begin{tabular}{|l|c|c|c|c|c|}
			\hline
			Model &  P & AIC & BIC &  $\Delta $AIC &  $\Delta $BIC \\
			\hline
			1. $\omega(k) = kc(1 + \lambda E)$ &0.524 & 2.4059 & 7.4430 & 0.0045 & 0.4799 \\ 
			\hline
			2. $\omega(k) = kc\sqrt{1 - 2\beta_0 E^2}$ & 0.516 & 2.4201 & 7.5127 & 0.0187 & 0.5697 \\ \hline
			3. $\omega(k) = kc\sqrt{1 - 2\alpha_0 E}$ & 0.526 & 2.4014 & 6.9639 & 0.0000 & 0.0000 \\
			\hline
		\end{tabular}
		\caption{Model comparison using p-value, AIC, and BIC criteria.}
		\label{coeff}
	\end{table}
	
	We observe that Model III provides the best statistical fit, having the lowest AIC and BIC values among the noncommutativity models. However, the relative differences $\Delta \mathrm{AIC}$ and $\Delta \mathrm{BIC} $ for Models I and II are small (less than $0.02$ and $< 0.58$, respectively), indicating that all three models perform comparably well in describing the observational data.
	
	In the low degree of freedom system we work in, while p-value determines model accuracy, the information criteria AIC and BIC provide better metrics for model selection. Based on Table~\ref{coeff}, we rank Model III as the best performing model, having the lowest AIC and BIC, followed by Model I and Model II. All models show statistically acceptable fits, having p-values of approximately 52\%, indicating that the models describe the data variance without signs of under- or overfitting.

\section{Discussions and final remarks} \label{Concl}

The possibility of the existence of a noncommutative structure of the spacetime, as well as the possibility of the existence of a minimum measurable length and of the related generalized uncertainty principle represent one of the most fundamental, and interesting problems in present day physics. Many methods have been proposed for the experimental/observational detection of these effects, including corrections to the Lamb shift, the Landau levels, and the tunneling current in a scanning tunneling microscope \cite{E1,E2},  the  
$^87$Rb  atom recoils experiments \cite{E3}, by using the properties of the gravitational wave events \cite{disp2,E4}, or the Stark effect in the hydrogen atom under an external electric field \cite{E5}. 
	
	In this work we analyzed the cosmological implications of three different spacetime non-commutativity models, each inducing a modified photon dispersion relation into the thermodynamic background. These models emerge from modified Heisenberg relations where the fundamental spacetime structure presents non-commuting coordinate operators, leading to modifications in the phase space, which results in particle propagation that deviate from the standard relativistic energy-momentum relation.

The first model is build on the commutation relations $\left[\hat{x}^i, \hat{t}\right] = i\lambda \hat{x}^i$, $\left[\hat{x}^i, \hat{x}^j\right] = 0$, which introduce a linear energy correction proportional to $\lambda \hbar\omega $, such that spacetime non-commutativity manifests as an additive term in the dispersion relation, $\omega(k) = kc(1+\lambda\hbar\omega)$. The second model presents a quadratic energy-dependent correction through the modified Heisenberg relations  $\left[\hat{x}^i, \hat{p}^j\right] = i\hbar \delta^{ij} (1 + \beta p^2)$ and  $\left[\hat{p}^i, \hat{p}^j\right] = 0$, which result in the deformed dispersion relation $\omega(k) = kc\sqrt{1-2\beta_0\hbar^2\omega^2}$, suggesting stronger non-commutative effects that become significant at higher energies. The third model incorporates the generalized Heisenberg relations $\left[\hat{x}^i, \hat{p}^j\right] = i\hbar \left(\delta^{ij} - \alpha_0(p \delta^{ij} + \frac{p^i p^j}{p}) + \alpha_0^2(p^2 \delta^{ij} + 3p^i p^j) \right)$ and  $ \left[\hat{p}^i, \hat{p}^j\right] = 0$, which induce a linear energy correction into the dispersion relation of the photon gas through $\omega(k) = kc\sqrt{1-2\alpha_0 \hbar \omega}$, representing an intermediate case between the other two deformation models considered.
	 
	Starting from the grand-canonical partition function, we derived the energy density and pressure for a photon gas with modified dispersion relations, obtaining analytical expressions in the low temperature limit relevant for Big Bang Nucleosynthesis. 
	These relations introduce temperature-dependent corrections to the standard radiation sector, scaling as $\lambda k_B T$, $\beta_0 (k_BT)^2$, and $\alpha_0 k_BT$ respectively, which affected the Hubble expansion rate during nucleosynthesis through their contribution to the plasma energy density.
	The altered thermodynamics relations impact the standard cosmological expansion timeline, eventually influencing the final abundance ratios of light elements formed in the primordial Universe.
	
	In order to estimate the deformation parameters of the three noncommutative spacetime models, we incorporated the thermodynamics for the modified dispersion relations of the photon gas into the \texttt{PRyMordial} codebase. The original python code was extended to include the additional contributions to the standard radiation energy density and pressure in the Hubble function evaluation, adding the relevant contributions into the New Physics sector.  We extended the \texttt{PRyMordial} BBN software into the \href{https://github.com/teomatei22/prynce/tree/main}{PRyNCe} program, which aims for a self-consistent evaluation of the cosmic expansion rate evolution, ensuring that the modified photon thermodynamics properly couples to the nuclear reaction network through the temperature-dependent $H(T)$ Hubble parameter. The nuclear reaction network, photon and neutrino temperature evolution, and freeze-out dynamics were solved numerically from $T = 10$ MeV to the keV scale.
	
	\begin{figure}[htbp!]
		\centering
		\includegraphics[scale = 0.28]{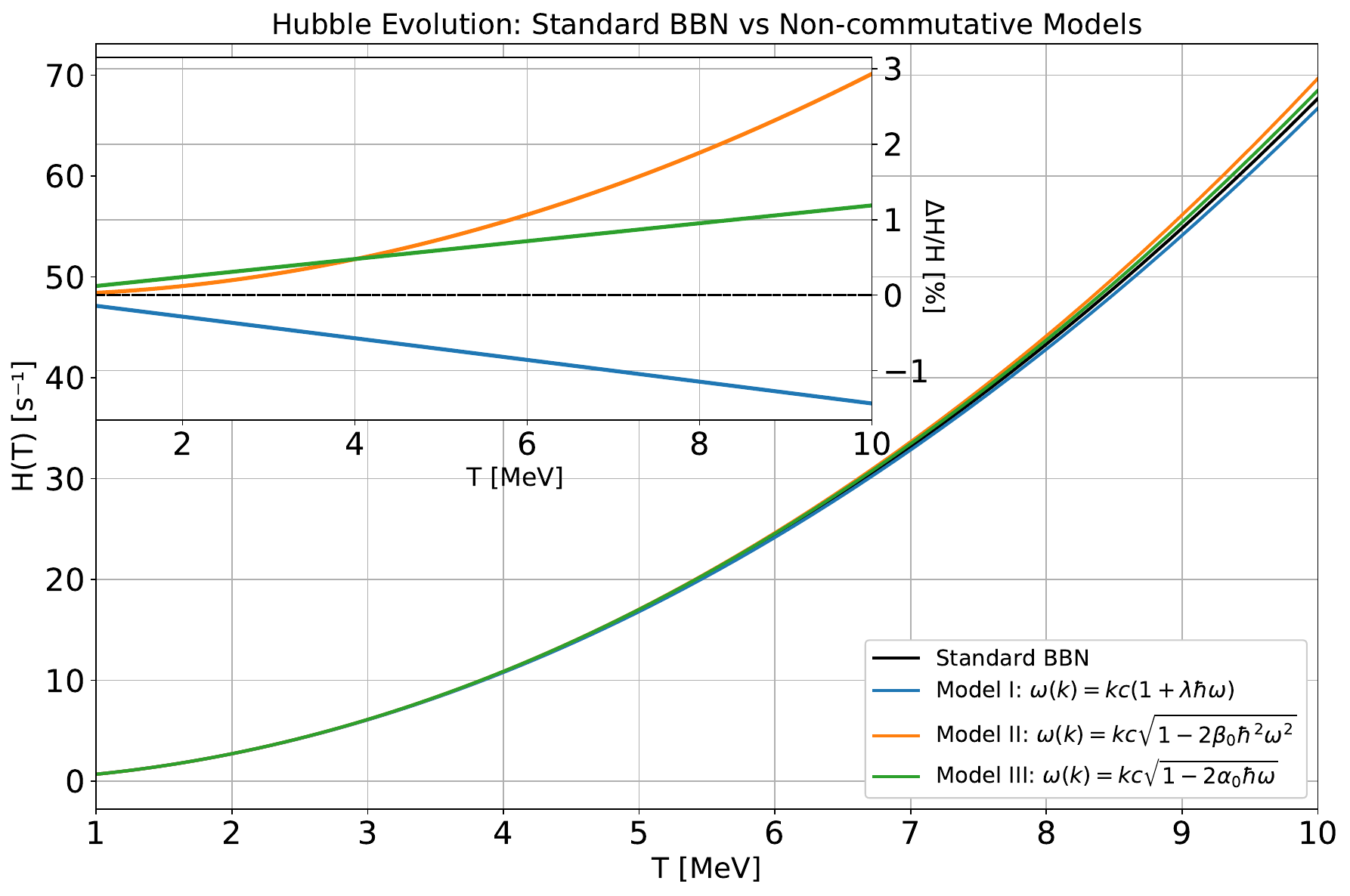}
		\caption{The temperature dependent Hubble function evolution for the modified dispersion models from spacetime noncommutativity and the Standard Model.}
		\label{HofT}
	\end{figure}
	
	 The discrepancy between the Standard Model physics and the noncommutative thermodynamics for the deformed photon gas can be traced back to the $t(T)$ function, which determines the evolution of all nucleosynthesis processes, including the evaluation of the modified Friedmann equation, and therefore $H(T)$. The Hubble parameter evolution is captured in Fig.~\ref{HofT}, which displays the behaviour of $H(T)$ function for all three dispersion relations considered, compared to the standard BBN scenario.
	 For the Hubble function evaluation we fixed the free parameters to their mean values and computed the thermal background for plasma evolution with and without the New Physics contribution.

	Each deformation model was subjected to an MCMC analysis and convergence was analyzed using the Gelman-Rubin statistic, with model parameters achieving $\hat{R} \lesssim 1.01$. Additionally, prior-posterior overlay analysis confirmed that the observational data effectively constrained the deformation parameters, with the posterior distributions lying within the uniform prior ranges imposed, while showing a clear convergence towards the smaller values from the original prior widths.

	The results show that all three deformation models compute light element abundances consistent with current observational bounds.  
	The best-fit values for the deformation parameter in the three noncommutativity spacetime models are 
	\begin{align}
	\lambda &= (2.91 \pm 2.10) \times 10^{-3} \; \rm{MeV}^{-1}, \notag \\
	\beta_0 &= (3.53 \pm 2.16) \times 10^{-4} \; \rm{MeV}^{-2},\notag \\
	\alpha_0 &= (8.55 \pm 4.73) \times 10^{-4}\; \rm{MeV}^{-1}. \notag 
	\end{align}
	
All free parameters satisfy the freeze-out constraint resulting from the deviations in the mass fraction estimates of the Helium-4 abundance, which require that the additional contribution to the energy density  at temperatures smaller than the freeze-out temperature remains below a value obtained from the $Y_p$ observations. Moreover, we impose that the equation of state parameter $w=\rho_{\text{deform}}/p_{\text{deform}}$ for the additional energy density and pressure coming from the deformed photon gas remains physical throughout the system's evolution.

 There are a large number of methods and approaches that have been proposed to constrain the noncommutativity parameters, and the generalized uncertainty relations.  For example, in \cite{disp1} it was suggested that the parameter $\lambda$ should satisfy the constraint $\lambda <10^{-3}$ GeV$^{-1}$, or $\lambda <10^{-6}$ MeV$^{-1}$. A stringent limit on the value of $\lambda$ can be obtained from ultrahigh energy cosmic rays experiments, or by quantum gravity arguments, which provide the constraints $\lambda <10^{-22}$ MeV$^{-1}$.  This bound is obtained by assuming a quantum gravity energy scale of the order of $E_{QG}>4\times 10^{18}$ GeV. 

As compared with these bounds, the BBN constraints for $\lambda$ indicate the possibility that in the early Universe noncommutative effects, via the modified photon dispersion relations, and the modified photon statistics,  may have played a more important role than, for example, in the case of the propagation of cosmic rays. 

Bounds on $\beta _0$ and $\alpha_0$ have been also obtained by using a large variety of observational and theoretical approaches \cite{disp2}. The results are usually expressed in terms of the dimensionless parameter $\beta =\beta_0 \times M_p^2c^4$, where $M_pc^2=1.22\times 10^{22}$ MeV is the Planck energy, with $M_p$ denoting the Planck mass, and $\beta$ is the dimensionless expression of the noncommutativity parameter. There is a large range of values that have been obtained for $\beta $, ranging from $10^{21}$ (electron tunneling), $10^{34}$ (electroweak measurements), $10^{36}$ (Lamb shift), $10^{50}$ (Landau levels), $10^{69}$ (perihelion precession Solar System), $10^{71}$ (perihelion precession Pulsar PRS B 1913+16 data), and $10^{78}$ (light deflection) (see \cite{disp2} and references therein). 

For the value of the dimensionless parameter $\beta $ as determined from the present investigation of the BBN we obtain the estimation $\beta_0 \times M_p^2c^4 \approx  3.53 \times 10^{-4} \times 1.488\times 10^{44}\approx 5.25\times 10^{40} $.  

This value of $\beta _0\times M_p^2c^4$ is very close to the value $10^{39}$ obtained from the $^{87}$Rb cold-atom-recoil experiments \cite{N1}. On the other hand, the upper bound $\beta _0\times M_p^2c^4<2.3\times 10^{60}$ was obtained in \cite{disp2} from the study of the data of gravitational waves event GW150914, which put some strong constraints on the difference $\Delta v$ between the speed of light and of the gravitational waves, $\left|\Delta v\right| \leq 10^{-17}$.  

The BBN estimations of $\beta _0\times M_p^2c^4$ obviously satisfies the gravitational wave constraints, however, they lead to a much stronger constraint on the parameter of the noncommutativity for the considered generalized uncertainty relation. On the other hand, there is a very big discrepancy between the large scale astrophysical and cosmological tests, and the local tests of the model-perihelion precession and light deflection,  performed at the level of the Solar System, or from the study of pulsars \cite{disp2}. 

There are also several  experimental bounds that have been obtained for $\alpha_0\times M_pc^2$, ranging from $10^8$ (anomalous magnetic moment of the muon), $10^{10}$ (Lamb shift), $10^{11}$ (electron tunneling), $10^{14}$ ($^{87}$Rb cold-atom-recoil experiments), $10^{17}$ (electroweak measurements and superconductivity), and $10^{23}$ (Landau levels), respectively (see \cite{disp2}, \cite{N1}, \cite{N2}, and references therein). The BBN constraints for the dimensionless value $\alpha_0\times M_pc^2$ is $\alpha_0\times M_pc^2\approx 1.043\times 10^{19}$. 

This value is very close to the bound $\alpha _0\times M_pc^2<1.8\times 10^{20}$ \cite{disp2}, obtained from the study of the gravitational waves emitted in the GW150914 event. 

On the other hand, the bounds on $\beta _0$ and $\alpha _0$ are different from those obtained from electroweak measurements, but still not fully incompatible with them. Hopefully, once more accurate measurements will be performed in the future, the differences in the upper bound on $\beta _0$  and $\alpha _0$ as predicted by the BBN estimations and elementary particle processes  will reduce significantly.

\begin{figure}[htbp!]
		\centering
		\includegraphics[scale=0.535]{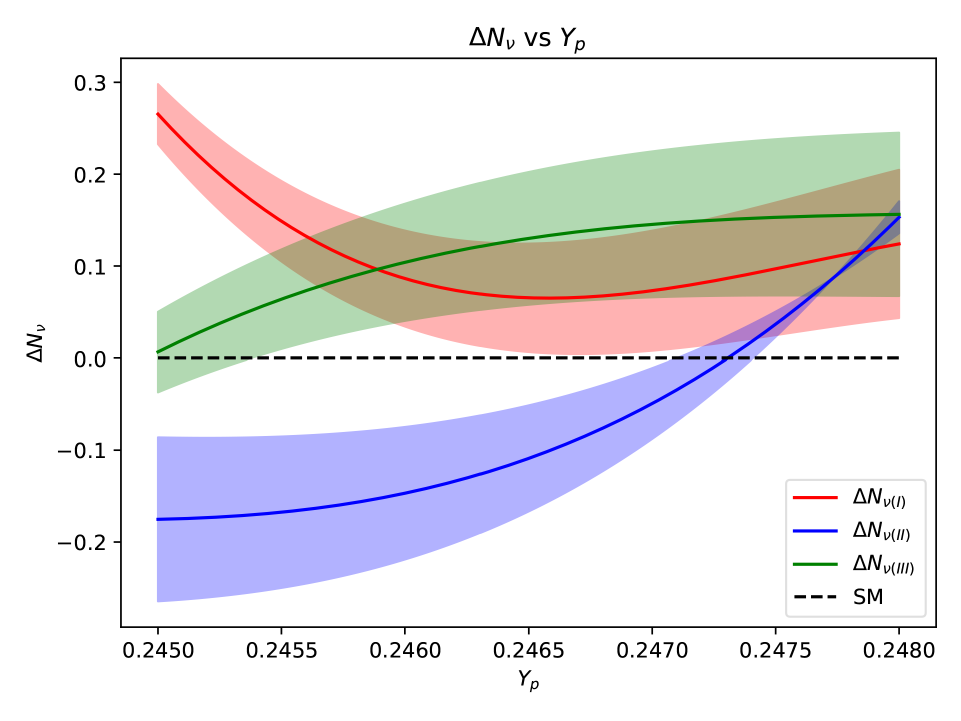}
		\caption{The deviation from relativistic species $\Delta N_\nu = N_{\mathrm{eff}} - 3.046$ for all three dispersion relations considered, represented against the corresponding $Y_p$ values calculated within \texttt{PRyNCe} framework.}
		\label{neff_yp}
	\end{figure}

	Several conclusions can be drawn from the current analysis of BBN quantities computed under modified photon dispersion relations. First, the effective number of relativistic degrees of freedom deviates from the Standard Model prediction by up to 0.295, including the 1$\sigma$ uncertainties, with all models satisfying the observational constraint $\Delta N_\nu < 0.3$,
	\begin{align}
		\Delta N_{\nu(I)} &= 0.128 \pm 0.098, \notag \\ 
		\Delta N_{\nu(II)} &= -0.187 \pm 0.108, \notag \\
		\Delta N_{\nu(III)} &= 0.162 \pm 0.093. \notag
	\end{align} 
	The correlation between $\Delta N_\nu$ and the helium mass fraction from our calculations is represented in Fig~\ref{neff_yp}, which clearly demonstrates how these deviations manifest. 
	
	The noncommutative spacetime relations modify the radiation energy density during the nucleosynthesis epoch, bringing for Model I a negative correlation between $\Delta N_\nu$ and $Y_p$, while Model II shows negative deviations from $N_{\text{eff}}$ which are positively correlated with $Y_p$ and Model III also presents a positive  $Y_p$ correlation, with positive deviations from the standard $N_{\text{eff}}$. 
	
	Second, the correlation structure among nuclear abundances remains similar across all three models, as shown in Table~\ref{table_correlations}. This consistency indicates that the underlying nuclear reaction networks dominate over the thermodynamic background evolution, so that each dispersion relation converges towards the standard BBN scenario at lower temperatures. While our results show broad agreement with SBBN data, notable deviations still exist, particularly for Lithium-7 production.
	
	Although the present research did not aim to solve the long-standing lithium problem, our calculations predict relevant constraints on $^7$Li under spacetime non-commutativity models, presented in Table~\ref{table_results}. The overproduction for Lithium-7 arises as we considered for computational optimization only 12 reactions in the nuclear network, with $t + \alpha \rightarrow ^7$Li + $\gamma$ and $^7$Li + $p \rightarrow \alpha + \alpha$ having a direct impact on its creation and destruction. This limitation impacts the overall lithium abundance through $N_{\text{eff}}$ and the baryon asymmetry parameter $\eta$. The latter is defined as  
	\begin{equation}
	\eta \equiv \frac{n_B}{n_\gamma}, \;\;
	n_\gamma = \frac{2\zeta(3)}{\pi^2}T_\gamma^3,
	\end{equation} 
	where $n_B$ is the baryon number density and $n_\gamma$ the equilibrium photon number density. With this value, the baryon density is written as  
	\begin{equation}
	\rho_B = m_p n_B = m_p \; \eta \; \eta_\gamma \propto \frac{\eta}{a^3},
	\end{equation}
	and directly determines the nuclear reaction rates during BBN. On the other hand, $\Delta N_{\nu}$ modifies the expansion rate by changing the total relativistic energy density,  
	\begin{equation}
	H^2 = \frac{\rho_{\text{tot}}}{3M_{\text{Pl}}^2}, \;\; 
	\rho_{\text{rad}} \propto \left( 1 + \frac{7}{8}\left(\frac{4}{11}\right)^{4/3} N_{\text{eff}} \right) T_\gamma^4 ,
	\end{equation}
	where $\rho_{\text{tot}} = \rho_{\text{rad}} + \rho_B + \rho_{\text{deform}}$,
	thus altering the time–temperature evolution that determines the freeze-out regime and nuclear production. The computational constraint for the reaction chain used was necessary as we focused on bounding small deformation parameters that arise from non-commutativity effects, which require precise observations of the primordial helium mass fraction $Y_p$ and deuterium-to-hydrogen ratio D/H for accurate constraint determination, rather than lithium abundance predictions.
	 
	  Our results remain significantly higher than the observed $^7$Li/H$=1.6\times 10^{-10}$ abundance, indicating that the modified dispersion relations do not dominate over the nuclear reaction effects implemented in \texttt{PRyMordial}, resulting in high production of lithium.

	Model comparison based on AIC and BIC shows that Model III is statistically preferred, having the lowest values for both criteria. However, the differences between the models are minimal, ($\Delta_{\mathrm{AIC}} < 0.2$ and $\Delta_{\mathrm{BIC}} < 0.57$), indicating that for such a low degree of freedom environment, all models perform nearly equally well in describing the nuclear abundances data. In terms of p-value statistics, the linear momentum deformation relation in Model III, $\omega(k) = kc \sqrt{1 - 2\alpha_0 \hbar \omega}$,
	corresponds to a p-value of 52.6\%, indicating a robust statistical fit to the data. The preference for Model III is reflected in Fig~\ref{HofT},  which clearly shows that the third dispersion model has the smallest deviation from the SBBN $H(T)$ evolution during the neutrino decoupling and BBN freeze-out epochs.
	
	These findings demonstrate that the deformed photon thermodynamics arising from space-time noncommutativity remain consistent with Standard BBN results, and place strong bounds on the magnitude of spacetime noncommutativity corrections during the nucleosynthesis process.  
	Future improvements in observational precision, in particular from upcoming CMB measurements and improved deuterium and helium observations, will enable placing even tighter constraints on the deformation parameters. These advances will provide important tests of early Universe physics, from modified gravity models to quantum gravity scenarios.

\end{document}